\documentclass[notitlepage,prd,showkeys,twocolumn,preprintnumbers,floatfix,superscriptaddress,nofootinbib]{revtex4-1}
\usepackage{amsmath}
\usepackage{amssymb}
\usepackage{xcolor}
\usepackage{bbm}
\usepackage{braket}
\usepackage{xspace}
\usepackage{mathtools}
\usepackage{graphicx}
\usepackage{enumerate}
\definecolor{darkblue}{rgb}{0.0,0.0,0.5}
\usepackage[colorlinks=true,
            linkcolor=darkblue,
            citecolor=darkblue,
            urlcolor=darkblue]{hyperref}
\usepackage{verbatim}
\usepackage{tabularx}
\usepackage{multirow}
\usepackage{enumitem}
\usepackage{orcidlink}
\usepackage[normalem]{ulem}

\newcommand{\SLJ}[3]{\ensuremath{{\:\!}^{{#1}\!}{#2}_{#3}}}

\usepackage{etoolbox}
\makeatletter
\patchcmd{\subsubsection}{\itshape}{\itshape\bfseries}{}{}
\makeatother

\newcommand{\Cc}{\mathcal{C}}
\newcommand{\Ec}{\mathcal{E}}
\newcommand{\Kc}{\mathcal{K}}
\newcommand{\Lc}{\mathcal{L}}
\newcommand{\Mc}{\mathcal{M}}
\newcommand{\Yc}{\mathcal{Y}}
\newcommand{\Rc}{\mathcal{R}}
\newcommand{\Fpv}{F_{\mathrm{pv}}}
\newcommand{\Kpar}{\text{K}}
\newcommand{\pcut}{p_{\rm cut}}

\newcommand{\svec}[1]{\mathbf{#1}}

\newcommand{\bbGamma}{{\mathpalette\makebbGamma\relax}}
\newcommand{\makebbGamma}[2]{%
  \raisebox{\depth}{\scalebox{1}[-1]{$\mathsurround=0pt#1\mathbb{L}$}}%
}

\newcommand{\lhcgamma}{\tilde{\gamma}}
\newcommand{\lhcinvkc}{\tilde{c}}

\newcolumntype{Y}{>{\centering\arraybackslash}X}
\newcolumntype{s}{>{\centering\arraybackslash\hsize=.2\hsize}X}

\definecolor{wm_green}{HTML}{115740}
\definecolor{wm_gold}{HTML}{B9975B}

\usepackage{mfirstuc}
\newcommand{\addReviewer}[2]{
\expandafter\newcommand\csname #1\endcsname[1]{{\sf \color{#2} {#1}:\,##1}}
\expandafter\newcommand\csname #1cor\endcsname[2]{{\color{#2} {#1}:\,\st{##1}{\sf ##2}}}
\expandafter\newcommand\csname #1color\endcsname{#2}
}

\usepackage{soul,color}

\definecolor{jlab_blue}{RGB}{47,122,121}
\definecolor{jlab_green}{RGB}{65,125,10}

\addReviewer{npl}{jlab_green}
\addReviewer{rb}{jlab_blue}

\newcommand{\ucb}{Department of Physics,
University of California,
Berkeley, CA 94720, USA}
\newcommand{\lbnl}{Nuclear Science Division,
Lawrence Berkeley National Laboratory, Berkeley,
CA 94720, USA}
\newcommand{\DAMTP}{DAMTP, University of Cambridge, Wilberforce Road, Cambridge, CB3 0WA, UK}
\begin{document}

\title{
Resolving the $T_{cc}^+$ with $\pi$ Exchange from Lattice QCD
}


\author{Nelson Pitanga Lachini~\orcidlink{0000-0003-1109-1473}}
\email[e-mail: ]{np612@cam.ac.uk}
\affiliation{\DAMTP}

\author{Ra\'ul A. Brice\~no~\orcidlink{0000-0003-1109-1473}}
\email[e-mail: ]{rbriceno@berkeley.edu}
\affiliation{\ucb}
\affiliation{\lbnl}

\collaboration{for the Hadron Spectrum Collaboration}

\begin{abstract}
We revisit the lattice QCD description of the doubly charmed tetraquark $T_{cc}^+$ using a finite-volume scattering formalism that incorporates one-pion exchange nonperturbatively, thereby making the nearest left-hand cut explicit while respecting unitarity and analyticity.
We simultaneously determine the $D D^\star \pi$ coupling and the isoscalar $DD^\star$ scattering amplitude by reanalyzing the previously computed finite-volume $DD^\star$ spectrum below the $D^\star D^\star$ threshold, on lattices with pion mass $m_\pi\approx391~{\rm MeV}$.
We find that the inclusion of pion exchange changes the $T_{cc}^+$ from a real-valued to a complex-valued pole in the second sheet of the $J^P=1^+$, $DD^\star$ amplitude below threshold.
We further predict that the $T_{cc}^+$ pole will move closer to the real-energy axis as the pion mass approaches its physical value.
\end{abstract}

\maketitle

\emph{\bf Introduction:}~The near-threshold $T_{cc}^+$~\cite{LHCb:2021vvq,LHCb:2021auc} provides an unusually sharp test of how long-range pion exchange and short-range $DD^\star$ dynamics combine to form this and perhaps other exotic states.
This has sparked numerous theoretical studies~\cite{Du:2021zzh,Meng:2022ozq,Du:2023hlu,Meng:2023bmz,Wang:2023iaz,Sun:2024wxz,Abolnikov:2024key,Dawid:2025wsn, Padmanath:2022cvl,Chen:2022vpo,Lyu:2023xro,Whyte:2024ihh,Collins:2024sfi,Prelovsek:2025vbr,Alharazin:2026lno,Stump:2025owq} aimed at confirming the existence of this state and understanding the role that pion exchange plays in the formation of the $T_{cc} ^+$, with the hope of ultimately drawing conclusions on the nature of this state.
This last point is particularly poignant given the fact that its quantum numbers require it to have two charm quarks, making it a good candidate for a tetraquark.
Meanwhile, its proximity to the $DD^\star$ threshold suggests it might be better described as a $DD^\star$ molecule.  

Ultimately, the nature of the $T_{cc}^+$ has to emerge from quantum chromodynamics (QCD).
Presently, the only nonperturbative method for studying QCD is lattice QCD.
Several lattice studies have investigated the $DD^\star$ dynamics in the $T_{cc}^+$~\cite{Padmanath:2022cvl,Chen:2022vpo,Lyu:2023xro,Whyte:2024ihh,Collins:2024sfi,Prelovsek:2025vbr,Alharazin:2026lno,Stump:2025owq}.
All of these have investigated this channel using different approaches, different values of the light-quark masses, and all draw the same conclusion: the $T_{cc}^+$ is indeed present.
This is already a major achievement, since experimental enhancements generally do not yield a smoking gun for a state in QCD.\footnote{A recent lattice QCD study of isoscalar $D\bar{D}$ found no evidence for near-threshold bound states or resonances~\cite{Wilson:2026bhu} in contrast to the structure reported by the Belle Collaboration~\cite{Belle:2017egg}, although the experimental evidence has not yet been independently confirmed.}

This has turned the community's attention to understanding the role of the pion exchange contribution.
This turns out to be quite critical, because, as was first pointed out in Refs.~\cite{Green:2021qol, Du:2023hlu, Dawid:2023jrj}, neglecting this contribution can invalidate conclusions drawn in previous lattice QCD studies.
Various efforts have been made to address this contribution~\cite{Meng:2021uhz,Bubna:2024izx,Raposo:2023oru,Dawid:2024oey,Yu:2025gzg,Raposo:2025dkb, Hansen:2024ffk}.\footnote{For implementations in studies of the $T_{cc}^+$ and other channels using lattice QCD data, see Refs.~\cite{Dawid:2025wsn,Rodas:2026zmm,Meng:2023bmz,Prelovsek:2025vbr}.} 

In this work, we focus on the framework laid out in Ref.~\cite{Raposo:2025dkb} in the reanalysis of the previously generated data presented in Ref.~\cite{Whyte:2024ihh}.
The reason to focus our attention on the formalism of Ref.~\cite{Raposo:2025dkb} is multifold.
It is currently the most general formalism for studying two-body dynamics directly from lattice QCD, and it accommodates effects from one-particle exchanges nonperturbatively.
Effects due to the left-hand cut (lhc) generated from the one-pion exchange (OPE) in $DD^\star$ scattering were ignored in the analysis of Ref.~\cite{Whyte:2024ihh}.
As recognized in that work, the proximity of the left-hand cut to the virtual bound state identified as the $T_{cc}^+$ could result in a large source of systematic uncertainty.
As we discuss in detail below, incorporating the OPE using the framework of Ref.~\cite{Raposo:2025dkb} not only improves the analysis, but it also affects previous findings regarding the $T_{cc}^+$.
We find strong evidence that the $T_{cc}^+$ is better described as a complex pole below threshold, instead of a real-valued virtual bound state.
Furthermore, the formalism of Ref.~\cite{Raposo:2025dkb} allows for a nonperturbative separation between the OPE and the remaining short-distance dynamics, encoded in a smooth matrix $\Kc_0$.
This separation allows a prediction of the evolution of the $T_{cc}^+$ as a function of the light-quark masses.

\begin{figure}[t]
    \centering
    \includegraphics[width=.49\textwidth]{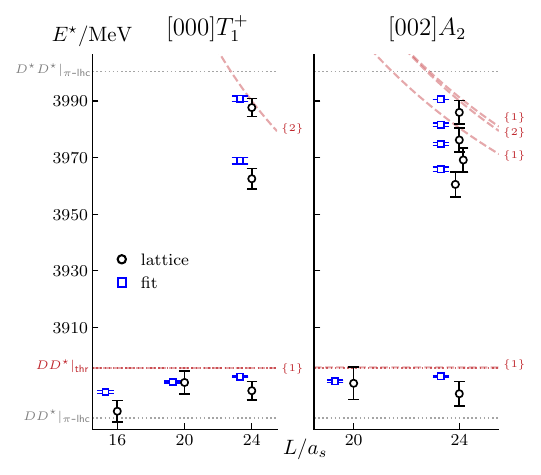}
     \caption{Finite-volume spectra in the rest-frame $[000]T^+_1$ (left) and moving-frame $[002]A_2$ (right) irreps.
     The black circles and blue squares are, respectively, the energies from the lattice and from the reference parametrization, both displaced horizontally for clarity.
     The $DD^\star$ threshold and single-pion left-hand branch points are indicated by horizontal dashed lines and dotted lines, respectively.
     The dashed curves represent the $DD^\star$ noninteracting energies with corresponding degeneracy given in curly brackets.
     The other irreps analyzed can be found in Sec.~\ref{sec:reflhc} of the Supplemental Material.}
    \label{fig:specvsL}
\end{figure}

\emph{\bf Review of formalism:}~Reference~\cite{Raposo:2025dkb} addressed the effects of the OPE without resorting to a low-energy effective field theory.
Instead, it begins by recognizing that for values of the quark masses for which the $D^\star$ is stable, the $T^+_{cc}$ appears as a pole of the $DD^\star$ scattering amplitude.
This amplitude can be expressed as a sum of two terms.
By projecting to a definite angular momentum $J$ and parity $P$, this amplitude can be written more explicitly as
\begin{align}
\Mc^{J^P} = \Mc^{J^P}_\Ec + \Mc^{J^P}_{\Kc_0} \,,
\label{eq:main_amp}
\end{align}
which combines the pure OPE ladder amplitude, $\Mc^{J^P}_{\mathcal E}$, with the short-distance contribution encoded in the two-body K matrix, $\Kc_0$. The latter is embedded within $\Mc^{J^P}_{\Kc_0}$, which we explicitly define in Sec.~\ref{sec:review} of the Supplemental Material.
Both $\Mc^{J^P}_\Ec $ and $ \Mc^{J^P}_{\Kc_0}$ should be understood as matrices over angular-momentum space, e.g. for $J^P=1^+$ they would be $2$ by $2$ matrices accommodating the coupled $\SLJ{3}{S}{1}$ and $\SLJ{3}{D}{1}$ waves.
Reference~\cite{Raposo:2025dkb} derived this identity using general principles of unitarity and analyticity, without assuming an effective Lagrangian.
The only generally unknown quantities are the matrix $\Kc_0$ and the $D D^\star \pi$ coupling $g$ entering the OPE.
Below, we discuss how these are constrained from lattice QCD observables.
For details on these relations, the definition of $\Mc_\Ec$ in terms of its integral equations, and the procedure used to calculate these functions, we refer the reader to Sec.~\ref{sec:amps} of the Supplemental Material.\footnote {There one also finds new expressions for the partial-wave-projected OPE for $P$ and $D$ waves that were not previously published.}
\\

In a finite volume, the physical scattering amplitude cannot be accessed directly.
Instead, one extracts the finite-volume energy spectrum.
Reference~\cite{Raposo:2025dkb} showed that, in the kinematic region where only $DD^\star$ can go on-shell, the finite-volume spectrum satisfies  
\begin{align}
\det_{Jm_J\ell}\left[
\left(\Kc_0\right)^{-1}
+ F - i \rho
+ \Cc_L
\right] = 0 \,,
\label{eq:main_QC}
\end{align}
where $\rho$ is the two-body phase-space factor, $F$ is the usual known finite-volume function~\cite{Briceno:2014oea}, and $\Cc_L$ contains the finite-volume effects induced by the OPE.
The $\Cc_L$ function depends explicitly on the $D D^\star \pi$ coupling $g$, and it vanishes in the limit $g\to0$.
We give a detailed description of this equation in Sec.~\ref{sec:review} of the Supplemental Material. 

\emph{\bf Spectrum analysis:}~Reference~\cite{Whyte:2024ihh} previously determined the finite-volume spectrum in the isoscalar channel with $DD^\star$ quantum numbers. 
Here, we analyze $36$ finite-volume energy levels below the $D^\star D^\star$ threshold to simultaneously constrain the coupling $g$ and the parametrizations of $\Kc_0$ using Eq.~\eqref{eq:main_QC}.
These energy levels were determined on anisotropic lattices with $2+1$ flavors of dynamical quarks and pion mass $m_\pi \approx 391$~\rm{MeV}, using spatial volumes with lengths $L/a_s = \{16, 20, 24\}$, where $a_s$ is the spatial lattice spacing~\cite{Edwards:2008ja,HadronSpectrum:2008xlg}.
The anisotropy, $\xi = a_s/a_t \approx 3.5$, the temporal lattice spacing, $a_t^{-1} \approx 5666~{\rm MeV}$, and the valence charm-quark tuning were previously determined in Refs.~\cite{Edwards:2012fx, HadronSpectrum:2012gic}.
At these quark masses, the $D$ and $D^\star$ mesons are stable and have masses of approximately $1886~{\rm MeV}$ and $2010~{\rm MeV}$, respectively.
Two-point correlation functions were computed using distillation~\cite{HadronSpectrum:2009krc}, and the finite-volume energies were extracted using the generalized eigenvalue method~\cite{Michael:1985ne,Luscher:1990ck,Blossier:2009kd,Dudek:2010wm}.
We refer the reader to Ref.~\cite{Whyte:2024ihh} for further details on the computational setup.

In general, the spectrum in a finite volume mixes different angular momentum and parity.
This is reflected in Eq.~\eqref{eq:main_QC} by the fact that the determinant is over angular momenta $J$, azimuthal components $m_J$, as well as orbital angular momenta $\ell$.
We address this by following a prescription already implemented in numerous lattice QCD calculations~\cite{Dudek:2012gj, PhysRevLett.129.252001, Lang:2025pjq, Woss:2018irj, Briceno:2017qmb, Wilson:2015dqa, BaryonScatteringBaSc:2023zvt}.
We begin by parametrizing the dominant partial-wave contributions to $\Kc_0$ in terms of real-valued functions of $s=E^{\star 2}$, where $E^{\star}$ is the energy in the center-of-momentum frame of the $DD^\star$ system.
We then constrain the coupling $g$ and the free $\Kc_0$ parameters by minimizing a $\chi^2$ function defined by the difference between the lattice energies and the energies predicted by Eq.~\eqref{eq:main_QC} in a correlated manner.\footnote{For details of the minimization procedure implemented here, we point the reader to Refs.~\cite{Wilson:2014cna,Woss:2020cmp}.}

In Fig.~\ref{fig:specvsL}, we show a subset of the spectra used in the analysis.
The figure shows the lattice QCD spectra as circle points for two irreps of the cubic group and its little group, defined by the total momentum, shown next to the irrep in units of $2\pi /L$.
The horizontal axis shows the spatial extent in lattice units.
Also shown are the energy levels in the limit of no interactions and their degeneracies, the $DD^\star$ threshold, and the left-hand branch point locations.
The square points show the result of the reference spectrum fit, defined by the value of $g = 12.57(33)$ and $4$ parameters describing the different components of $\Kc_0$.
This was the result of a chi-squared minimization resulting in $\chi^2/N_\mathrm{dof} = 27.98/31 = 0.90$, where $N_\mathrm{dof}$ is the number of degrees of freedom.
This is the statistically preferred fit according to the Akaike information criterion (AIC) applied across all explored $\Kc_0$ parametrizations.
Details of this, including information on all variations obtained, can be found in Sec.~\ref{sec:specanalysis} of the Supplemental Material.

A significant conclusion of the analysis presented here is that including the effects of the OPE allows the spectrum to be described with fewer parameters while improving the fit quality.
In particular, the fit improves from $\chi^2 / N_\mathrm{dof} \approx 1.7$ when using $7$ parameters, down to $\chi^2 / N_\mathrm{dof} \approx 1$ using $3$ -- $5$ parameters (see Secs.~\ref{sec:refluscher} and~\ref{sec:par_vars} of the Supplemental Material).
By averaging over all parametrizations and including their systematic deviations following the AIC-based prescription from Refs.~\cite{Jay:2020jkz,Pefkou:2021fni}, we obtain $g=11.8(7)$.

Another important observation is that even though in Ref.~\cite{Whyte:2024ihh} it was necessary to include both $J^P = 0^-$ and $2^-$ $P$-wave contributions to the K matrix, here we find that only a nonzero $\SLJ{3}{P}{2}$ term in $\Kc_0$ is necessary to well describe the spectrum.
On the other hand, the $0^-$ contribution to the scheme-dependent $\Kc_0$ is consistent with zero and, as a consequence, the corresponding scattering amplitude is purely generated by OPE contributions via $\Mc^{J^P}_{\mathcal E}$ in Eq.~\eqref{eq:main_amp}.
\\

\emph{\bf Amplitude analysis:}~With the $g$ and $\Kc_0$ determined, we use Eq.~\eqref{eq:main_amp} to reconstruct the coupled $\SLJ{3}{S}{1}$–$\SLJ{3}{D}{1}$, $DD^\star$ amplitude in $J^P=1^+$, where the $T_{cc}^+$ resides, as well as the $J^P= 0^-$ and $2^-$, $P$-wave amplitudes. 
Most of the ingredients needed for obtaining the partial-wave projected OPE and solving the $\Mc_\Ec$ integral equations can be adopted directly from existing literature~\cite{Jackura:2023qtp, Raposo:2025dkb, Dawid:2023jrj}, but we present all details, including some not previously presented in the literature, in Sec.~\ref{sec:amps} of the Supplemental Material. 

\begin{figure}[t]
    \centering
    \includegraphics[width=.45\textwidth]{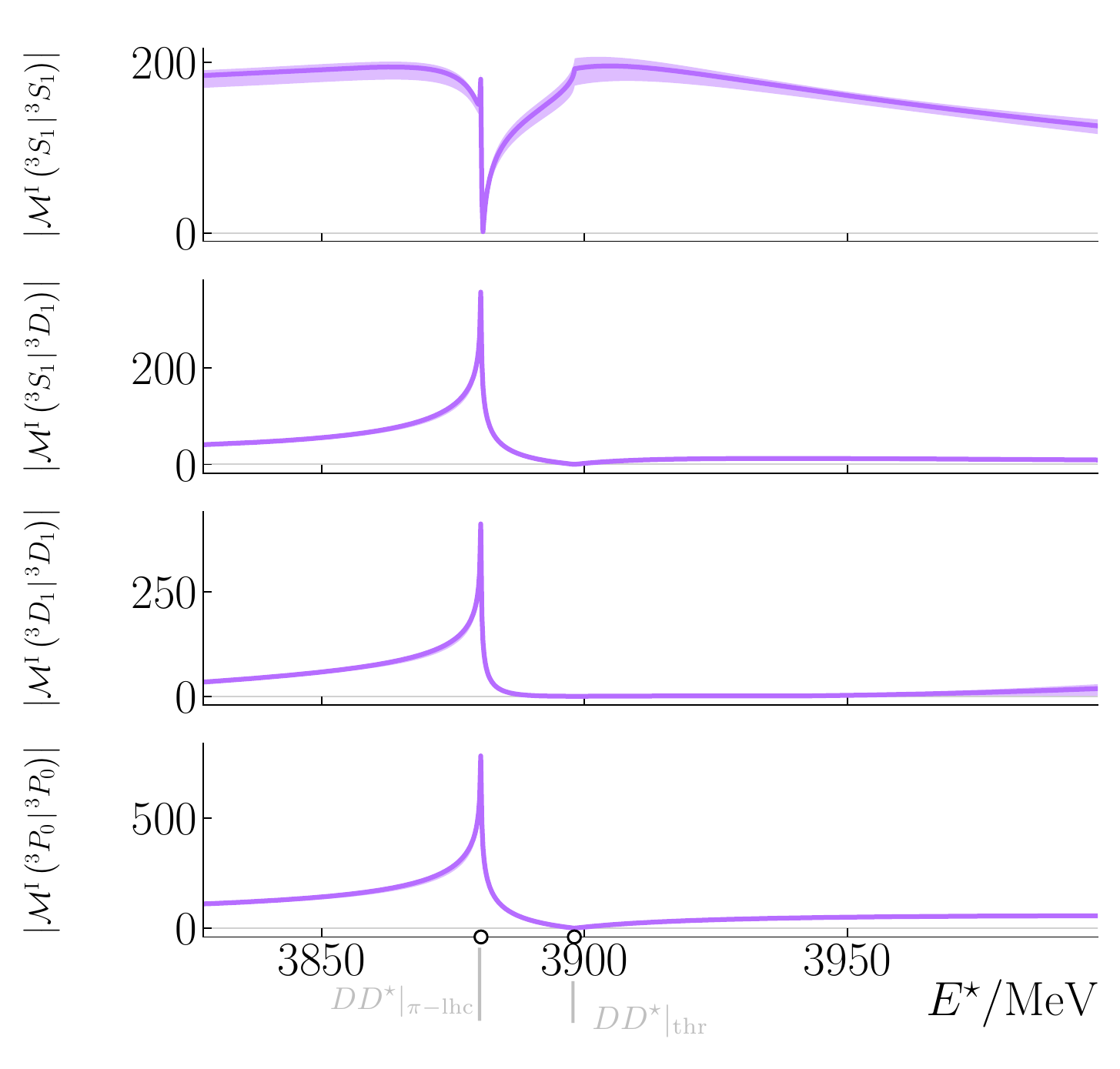}
     \caption{ 
     Absolute value of the physical-sheet isoscalar $DD^\star$ scattering amplitudes obtained from the reference parametrization.
     The first three panels show the coupled $J^P=1^+$ in the $\SLJ{3}{S}{1}$–$\SLJ{3}{D}{1}$ basis, and the fourth the $J^P=0^-$ $\SLJ{3}{P}{0}$ amplitude.}
    \label{fig:Tcc_M_abs}
\end{figure}

The calculation includes the coupled $\SLJ{3}{S}{1}$ and $\SLJ{3}{D}{1}$ partial waves, and the logarithmic structure of the OPE enters through the Legendre function, given explicitly in Eq.~\eqref{eq:Ec_pw}.
This makes the left-hand branch point explicit in the amplitudes.
The first-sheet amplitudes for the previously mentioned reference parametrization of $\Kc_0$ are given in Fig.~\ref{fig:Tcc_M_abs}.
For all amplitudes, one sees the canonical logarithmic divergence due to the presence of the left-hand cut.
In Sec.~\ref{sec:scatamps_poles} of the Supplemental Material, we present the amplitudes for all parametrizations considered, including for the $\SLJ{3}{P}{2}$ channel, and we also show evidence that all amplitudes satisfy the unitarity bound above threshold.

\begin{figure}[t]
    \centering
    \includegraphics[width=.45\textwidth]{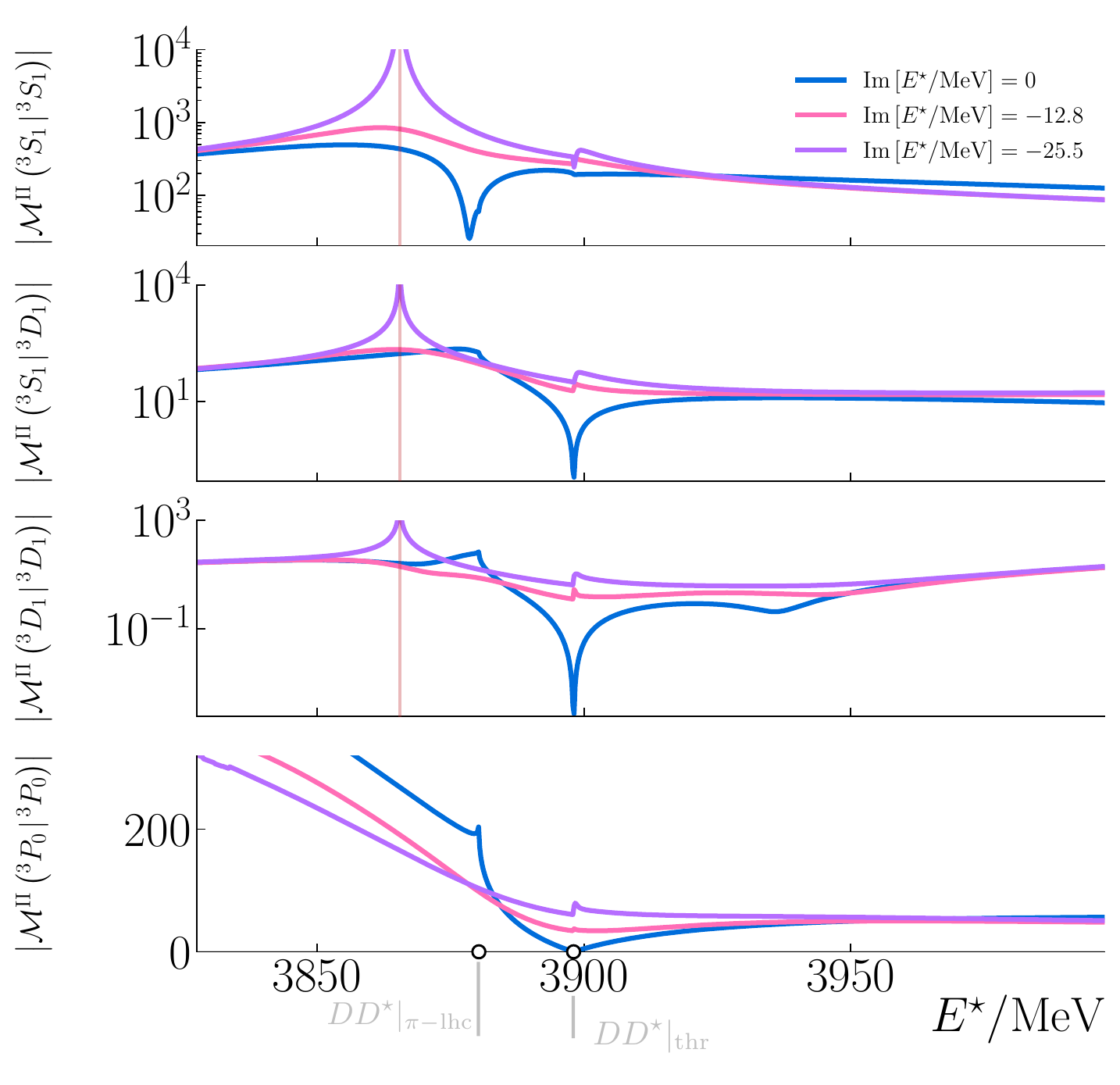}
     \caption{As in Fig.~\ref{fig:Tcc_M_abs}, but where the central value of the second-sheet amplitudes are evaluated along complex-energy trajectories for the reference parametrization.
     The vertical lines indicate the real part of the $J^P=1^+$ pole found.}
    \label{fig:Tcc_M_sheetII_complex_abs}
\end{figure}

Given the amplitude on the first sheet and real values of energy, it is possible to analytically continue this to the second Riemann sheet using
\begin{align}
\Mc^{\rm II}
=
\left[
\left(\Mc^{\rm I}\right)^{-1}
+2 i \rho
\right]^{-1} \,.
\label{eq:sheetII}
\end{align}
The resulting analytic continuation for the $J^P=1^+$ and $0^-$ amplitudes into the second sheet is shown in Fig.~\ref{fig:Tcc_M_sheetII_complex_abs}.
As can be seen from the figure, when the imaginary part of the energy is exactly equal to zero, none of the amplitudes acquire a pole.
This is in direct contrast with Refs.~\cite{Padmanath:2022cvl, Whyte:2024ihh}, where a virtual bound state pole was found. 

Analytically continuing the formalism used here for complex energies requires care.
This is because we need to solve integral equations involving two singular kernels, the OPE and the two-particle propagator, in the integrand.
This class of integral equations and their analytic continuation was recently discussed in Ref.~\cite{Dawid:2023jrj}.
We summarize the key ingredients in our implementation in Sec.~\ref{sec:scatamps_poles} of the Supplemental Material.
The main takeaway message is that for the parametrizations considered, we verify that the results quoted below lie within the domain of analyticity of the integral equations.

\begin{figure}[t]
    \centering
    \includegraphics[width=.48\textwidth]{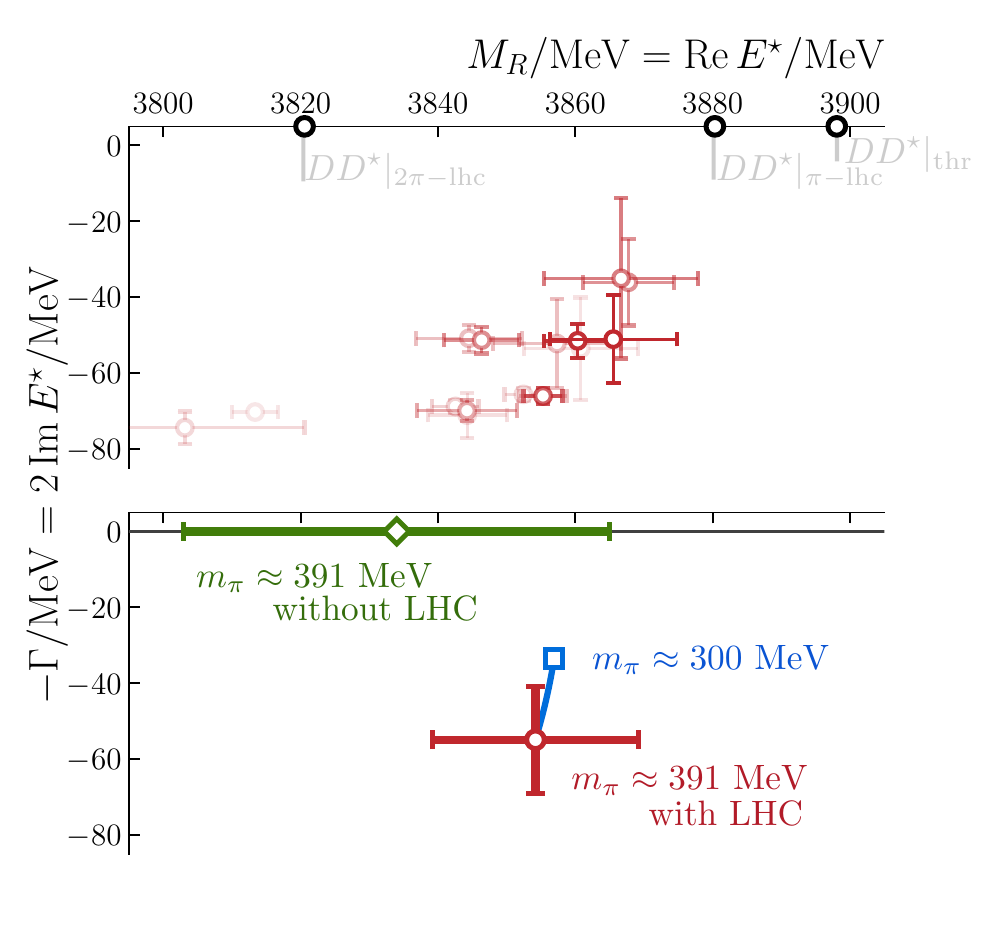}
     \caption{
     Upper panel: red circles represent the locations of the nearest $T_{cc}^+$ pole in the complex plane, using $E_{\rm pole } = M_R-i\Gamma/2$, for all parametrizations considered.
     The opacity of each point is set by its relative AIC weight as discussed in Sec.~\ref{sec:specanalysis} of the Supplemental Material.
     Lower panel: in blue is shown the pion-mass evolution of the pole location down to $m_\pi \approx 300$~{\rm MeV}, as discussed in the text.
     The red circle is the AIC-weighted average of the $T_{cc}^+$ pole over all variations, and the green diamond is the result neglecting the left-hand cut~\cite{Whyte:2024ihh}.
     Also shown are the locations of the $DD^\star$ threshold and the single- and two-pion left-hand branch points for $m_\pi \approx 391$~{\rm MeV}.
     }
    \label{fig:pole}
\end{figure}

We find the $T_{cc}^+$ to be a complex-valued pole in the second Riemann sheet of the $J^P=1^+$ amplitude.
We verify that this pole is present on all the explored parametrizations of the $\Kc_0$ matrix.
In the upper panel of Fig.~\ref{fig:pole}, we summarize the pole positions coming from all such variations, with the opacity of each point being proportional to the corresponding AIC weight, normalized to the largest weight. 

Using the same AIC prescription used above to obtain an average value for the exchange coupling $g$, we find that the pole and residue couplings to the $1^+$ channels are
\begin{align}
 E_{\rm pole}
&=
\bigg(3854(15)-\frac{i}{2}55(14)\bigg)~{\rm MeV} \,,
\\
c(\SLJ{3}{S}{1} )&=
11(2)\,e^{i\,0.98(3)\frac{\pi}{2}}\times 10^3~{\rm MeV} \,,
\\
c(\SLJ{3}{D}{1})&=
1.8(3)\,e^{-i\,1.0(2)\frac{\pi}{2}}\times 10^3~{\rm MeV} \,.
\end{align}
Equating this to $M_R-i\Gamma/2$, we find that, at this value of the light-quark masses, the $T_{cc}^+$ has a ``width'' of $\Gamma= 54(14)~\rm MeV$.\footnote{Because the real part of the pole lies below the $DD^\star$ threshold, this width cannot be understood as the standard decay width.}
In the lower panel, our final averaged pole position is compared to the real-valued pole reported in Ref.~\cite{Whyte:2024ihh}.
To further illustrate this, in Fig.~\ref{fig:Tcc_M_sheetII_complex_abs}, we present slices of the amplitudes in the second Riemann sheet for complex energies for our reference parametrization.
We see that when the imaginary part of the energy approaches the pole, all components of the $J^P=1^+$ amplitude indeed diverge.
In contrast, we do not see evidence for a pole in the $\SLJ{3}{P}{0}$ amplitude.\footnote{Reference~\cite{Whyte:2024ihh} reported a $J^P=0^-$, $P$-wave resonance pole inconsistently found across different K-matrix parametrizations, such that no definitive conclusion was drawn.}

Although we cannot yet predict the dependence of either $\Kc_0$ or $g$ on the light-quark mass, the pion mass $m_\pi$ is expected to vary most rapidly as the light-quark mass is changed.
With this in mind, we explore the light-quark mass dependence of the pole by varying $m_\pi$, which appears explicitly in the definition of the OPE.
The blue trajectory ending at the $m_\pi \approx 300$~{\rm MeV} is obtained by changing only the exchanged pion mass in the OPE kernel while holding the $\Kc_0$ parameters and $g$ fixed to their reference values.
In summary, we find that the decay width should decrease as one approaches quark masses corresponding to the physical pion mass.
\\

\emph{\bf Conclusion and outlook:}~We have presented the first lattice QCD computation of the $T_{cc}^+$ pole that includes one-pion exchange nonperturbatively by simultaneously determining the $DD^\star\pi$ coupling and the $DD^\star$ scattering amplitude solely from a large set of finite-volume energies, with good statistical precision.
We find that the inclusion of the OPE significantly improves the statistical analysis of the finite-volume spectrum, and it changes the interpretation of the $T_{cc}^+$ pole.

It is useful to compare the findings in this work with previous hybrid analyses of the $T_{cc}^+$ state based on fewer lattice QCD energy levels.
In Ref.~\cite{Dawid:2025wsn}, a mix of different inputs was used to estimate the pole position of the $T_{cc}^+$ at ${m_\pi\approx280~{\rm MeV}}$.
The primary input was previously generated lattice QCD spectra generated in Ref.~\cite{Collins:2024sfi}. 
The second set of inputs was the Heavy Meson Chiral Perturbation Theory (HM$\chi$PT) parameter $f_\pi=130~{\rm MeV}$ with $g_0\approx 0.5$ being constrained by lattice QCD form factors~\cite{Becirevic:2012pf}, corresponding to $g_{DD^\star\pi}=2g_0\sqrt{m_Dm_{D^\star}}/f_\pi \approx 15.5$
, and related to the coupling used here via
${g= \sqrt{3/2}\,g_{DD^\star\pi} \approx 19}$.
These were used to supplement the insufficient constraints from the lattice QCD data to isolate the OPE coupling.
We see this prediction to have the same order of magnitude as our determination of the coupling, which is reassuring because although the definition of $\Kc_0$ is scheme dependent, the $g$ coupling is not. 
Reference~\cite{Dawid:2025wsn} also found that for $m_\pi\approx280~{\rm MeV}$ the $T_{cc}^+$ was identified as a subthreshold second-sheet pole with $\Gamma \approx 30~{\rm MeV}$, which is consistent with our predicted trend shown in Fig.~\ref{fig:pole}, namely that if we assume our fitted parameters do not vary by lowering the exchanged pion mass, the pole moves toward the real-energy axis.
Other previous studies employing effective approaches to analyze lattice QCD data supplemented by external constraints reached similar conclusions~\cite{Du:2023hlu,Meng:2023bmz,Collins:2024sfi,Prelovsek:2025vbr}. This level of agreement is notable given that our results are based only on the lattice QCD spectrum, making no use of external inputs. 

We close by going back to three major goals that the theoretical community has laid out for itself.
First, it is now clear that the $T_{cc}^+$ is indeed a state of QCD.
Second, as previously pointed out~\cite{Du:2023hlu, Dawid:2023jrj}, the left-hand cut due to one-pion exchange can strongly modify the conclusion drawn from lattice QCD calculations, and in this work we have shown that the inclusion of the left-hand cut in the analysis changed the previous observation of a virtual bound state in Ref.~\cite{Whyte:2024ihh}, to a subthreshold complex pole.
As a result, it is now clear that analysis of lattice QCD spectra in the vicinity of a left-hand cut that does not explicitly take into account the impact of the left-hand cut must be taken with a grain of salt.\footnote{Ref.~\cite{Dawid:2023jrj} did find that real bound-state poles are largely unaffected by left-hand cuts.}
Furthermore, this study shows that it is indeed possible to constrain all required parameters, including the $D D^\star \pi$ coupling, directly from lattice QCD spectrum without having to resort to effective field theories, such as HM$\chi$PT, which may have an overly restricted region of applicability.

Finally, the third goal of identifying the nature of the $T_{cc}^+$ is much harder to achieve.
Spectroscopy and models alone are unlikely to draw definitive conclusions as to whether this is best described as a compact tetraquark or a large molecular configuration.
At this point, the internal structure of this state will likely need to be probed using local currents, as is already being done for simple QCD states~\cite{Pefkou:2021fni,HadStruc:2021qdf,HadStruc:2024rix,Lin:2021brq,Alexandrou:2019ali,Bhattacharya:2023ays,Bhattacharya:2024wtg,Batelaan:2025vhx,Batelaan:2025vbb,Shultz:2015pfa}.
Efforts are already underway for studying the structure of scattering states via lattice QCD~\cite{Briceno:2015tza,Baroni:2018iau,Bernard:2012bi,Meissner:2026zos}, and the relationship between the new classes of amplitudes that emerge and form factors of composite states has been derived~\cite{Briceno:2020vgp} and tested~\cite{Briceno:2019nns,Moscoso:2026wmz}.
Until recently, these formalisms have ignored the possible left-hand cut effects, but efforts are underway to include these nonperturbatively~\cite{Raposo:2026}. 

With the progress presented here and the formal developments underway, it is reasonable to have a definitive resolution of the nature of the $T_{cc}^+$ directly from QCD in the years to come.
Finally, given that the program being developed is independent of an effective field theory and is generally applicable, we can expect these principles and tools to be transferable to many other channels, including the two-nucleon sector~\cite{BaryonScattering:2025ziz, Horz:2020zvv, Amarasinghe:2021lqa, Detmold:2024iwz}.
\\

\section*{Acknowledgments}
We thank our colleagues within the Hadron Spectrum Collaboration (www.hadspec.org) for useful discussions, and would particularly like to thank J.~Dudek, M. T. Hansen, Felipe G. Ortega-Gama, David J. Wilson, Christopher E. Thomas, and A. W. Jackura for helpful comments on a previous version of this manuscript. The authors also thank André Baião Raposo, Caroline S. R. Costa, and Sebastian M. Dawid for useful discussions.
NPL acknowledges support from the U.K. Science and Technology Facilities Council (STFC) [grant numbers ST/T000694/1, ST/X000664/1].
RAB was partly supported by the U.S. Department of Energy, Office of Science, Office of Nuclear Physics under Award No. DE-SC0025665 and No. DE-AC02-05CH11231.

\bibliography{ref}

@article{Meissner:2026zos,
    author = "Mei{\ss}ner, Ulf-G. and Rusetsky, Akaki and Sakthivasan, Ajay S. and Schierholz, Gerrit and Wu, Jia-Jun",
    title = "{Form factors of the {\ensuremath{\rho}} meson from effective field theory and the lattice}",
    eprint = "2602.23044",
    archivePrefix = "arXiv",
    primaryClass = "hep-lat",
    reportNumber = "DESY-26-030",
    doi = "10.1007/JHEP06(2026)164",
    journal = "JHEP",
    volume = "06",
    pages = "164",
    year = "2026"
}

@article{Bernard:2012bi,
    author = "Bernard, V. and Hoja, D. and Meissner, U. G. and Rusetsky, A.",
    title = "{Matrix elements of unstable states}",
    eprint = "1205.4642",
    archivePrefix = "arXiv",
    primaryClass = "hep-lat",
    doi = "10.1007/JHEP09(2012)023",
    journal = "JHEP",
    volume = "09",
    pages = "023",
    year = "2012"
}

@article{Belle:2017egg,
    author = "Chilikin, K. and others",
    collaboration = "Belle",
    title = "{Observation of an alternative $\chi_{c0}(2P)$ candidate in $e^+ e^- \rightarrow J/\psi D \bar{D}$}",
    eprint = "1704.01872",
    archivePrefix = "arXiv",
    primaryClass = "hep-ex",
    reportNumber = "BELLE-PREPRINT-2017-07, KEK-PREPRINT-2017-3",
    doi = "10.1103/PhysRevD.95.112003",
    journal = "Phys. Rev. D",
    volume = "95",
    pages = "112003",
    year = "2017"
}

@article{Wilson:2015dqa,
    author = "Wilson, David J. and Briceno, Raul A. and Dudek, Jozef J. and Edwards, Robert G. and Thomas, Christopher E.",
    title = "{Coupled $\pi\pi, K\bar{K}$ scattering in $P$-wave and the $\rho$ resonance from lattice QCD}",
    eprint = "1507.02599",
    archivePrefix = "arXiv",
    primaryClass = "hep-ph",
    reportNumber = "DAMTP-2015-34, JLAB-THY-15-2101",
    doi = "10.1103/PhysRevD.92.094502",
    journal = "Phys. Rev. D",
    volume = "92",
    number = "9",
    pages = "094502",
    year = "2015"
}

@article{BaryonScatteringBaSc:2023zvt,
    author = "Bulava, John and others",
    collaboration = "Baryon Scattering (BaSc)",
    title = "{Two-Pole Nature of the {\ensuremath{\Lambda}}(1405) resonance from Lattice QCD}",
    eprint = "2307.10413",
    archivePrefix = "arXiv",
    primaryClass = "hep-lat",
    reportNumber = "MIT-CTP/5579",
    doi = "10.1103/PhysRevLett.132.051901",
    journal = "Phys. Rev. Lett.",
    volume = "132",
    number = "5",
    pages = "051901",
    year = "2024"
}

@article{Briceno:2017qmb,
    author = "Briceno, Raul A. and Dudek, Jozef J. and Edwards, Robert G. and Wilson, David J.",
    title = "{Isoscalar $\pi\pi, K\overline{K}, \eta\eta$ scattering and the $\sigma, f_0, f_2$ mesons from QCD}",
    eprint = "1708.06667",
    archivePrefix = "arXiv",
    primaryClass = "hep-lat",
    reportNumber = "JLAB-THY-17-2534",
    doi = "10.1103/PhysRevD.97.054513",
    journal = "Phys. Rev. D",
    volume = "97",
    number = "5",
    pages = "054513",
    year = "2018"
}

@article{HadStruc:2021qdf,
    author = "Egerer, Colin and others",
    collaboration = "HadStruc",
    title = "{Transversity parton distribution function of the nucleon using the pseudodistribution approach}",
    eprint = "2111.01808",
    archivePrefix = "arXiv",
    primaryClass = "hep-lat",
    reportNumber = "JLAB-THY-21-3521",
    doi = "10.1103/PhysRevD.105.034507",
    journal = "Phys. Rev. D",
    volume = "105",
    number = "3",
    pages = "034507",
    year = "2022"
}

@article{HadStruc:2024rix,
    author = "Dutrieux, Herv{\'e} and Edwards, Robert G. and Egerer, Colin and Karpie, Joseph and Monahan, Christopher and Orginos, Kostas and Radyushkin, Anatoly and Richards, David and Romero, Eloy and Zafeiropoulos, Savvas",
    collaboration = "HadStruc",
    title = "{Towards unpolarized GPDs from pseudo-distributions}",
    eprint = "2405.10304",
    archivePrefix = "arXiv",
    primaryClass = "hep-lat",
    reportNumber = "JLAB-THY-24-4059, JLAB-THY-24-4059",
    doi = "10.1007/JHEP08(2024)162",
    journal = "JHEP",
    volume = "08",
    pages = "162",
    year = "2024"
}

@article{Lin:2021brq,
    author = "Lin, Huey-Wen",
    title = "{Nucleon helicity generalized parton distribution at physical pion mass from lattice QCD}",
    eprint = "2112.07519",
    archivePrefix = "arXiv",
    primaryClass = "hep-lat",
    reportNumber = "MSUHEP-21-024",
    doi = "10.1016/j.physletb.2021.136821",
    journal = "Phys. Lett. B",
    volume = "824",
    pages = "136821",
    year = "2022"
}

@article{Alexandrou:2019ali,
    author = "Alexandrou, C. and others",
    title = "{Moments of nucleon generalized parton distributions from lattice QCD simulations at physical pion mass}",
    eprint = "1908.10706",
    archivePrefix = "arXiv",
    primaryClass = "hep-lat",
    doi = "10.1103/PhysRevD.101.034519",
    journal = "Phys. Rev. D",
    volume = "101",
    number = "3",
    pages = "034519",
    year = "2020"
}

@article{Bhattacharya:2023ays,
    author = "Bhattacharya, Shohini and Cichy, Krzysztof and Constantinou, Martha and Gao, Xiang and Metz, Andreas and Miller, Joshua and Mukherjee, Swagato and Petreczky, Peter and Steffens, Fernanda and Zhao, Yong",
    title = "{Moments of proton GPDs from the OPE of nonlocal quark bilinears up to NNLO}",
    eprint = "2305.11117",
    archivePrefix = "arXiv",
    primaryClass = "hep-lat",
    doi = "10.1103/PhysRevD.108.014507",
    journal = "Phys. Rev. D",
    volume = "108",
    number = "1",
    pages = "014507",
    year = "2023"
}

@article{Bhattacharya:2024wtg,
    author = "Bhattacharya, Shohini and Cichy, Krzysztof and Constantinou, Martha and Gao, Xiang and Metz, Andreas and Miller, Joshua and Mukherjee, Swagato and Petreczky, Peter and Steffens, Fernanda and Zhao, Yong",
    title = "{Moments of axial-vector GPD from lattice QCD: quark helicity, orbital angular momentum, and spin-orbit correlation}",
    eprint = "2410.03539",
    archivePrefix = "arXiv",
    primaryClass = "hep-lat",
    reportNumber = "LA-UR-24-29020",
    doi = "10.1007/JHEP01(2025)146",
    journal = "JHEP",
    volume = "01",
    pages = "146",
    year = "2025"
}

@article{Batelaan:2025vhx,
    author = "Batelaan, Mischa and Dudek, Jozef J. and Edwards, Robert G.",
    collaboration = "Hadron Spectrum",
    title = "{{\ensuremath{\eta}} and {\ensuremath{\eta}}' Production in J/{\ensuremath{\psi}} Radiative Decays from Quantum Chromodynamics}",
    eprint = "2506.09306",
    archivePrefix = "arXiv",
    primaryClass = "hep-lat",
    reportNumber = "JLAB-THY-25-4368",
    doi = "10.1103/tdk2-bk8w",
    journal = "Phys. Rev. Lett.",
    volume = "135",
    number = "16",
    pages = "161904",
    year = "2025"
}

@article{Batelaan:2025vbb,
    author = "Batelaan, Mischa and Dudek, Jozef J. and Edwards, Robert G.",
    collaboration = "Hadron Spectrum",
    title = "{{\ensuremath{\eta}} and {\ensuremath{\eta}}' meson production in J/{\ensuremath{\psi}} radiative decays from lattice QCD}",
    eprint = "2506.09305",
    archivePrefix = "arXiv",
    primaryClass = "hep-lat",
    reportNumber = "JLAB-THY-25-4370",
    doi = "10.1103/jvpg-3ph4",
    journal = "Phys. Rev. D",
    volume = "112",
    number = "7",
    pages = "074505",
    year = "2025"
}

@article{Shultz:2015pfa,
    author = "Shultz, Christian J. and Dudek, Jozef J. and Edwards, Robert G.",
    title = "{Excited meson radiative transitions from lattice QCD using variationally optimized operators}",
    eprint = "1501.07457",
    archivePrefix = "arXiv",
    primaryClass = "hep-lat",
    reportNumber = "JLAB-THY-15-2004",
    doi = "10.1103/PhysRevD.91.114501",
    journal = "Phys. Rev. D",
    volume = "91",
    number = "11",
    pages = "114501",
    year = "2015"
}

@unpublished{Raposo:2026,
  author = {Bai{\~a}o-Raposo, Andr{\'e} and Brice{\~n}o, Ra{\'u}l A. and Lang, Nicolas and Ortega-Gama, Felipe G. and Pol, Bianca},
  year   = {2026},
  note   = {In preparation}
}

@article{Jay:2020jkz,
    author = "Jay, William I. and Neil, Ethan T.",
    title = "{Bayesian model averaging for analysis of lattice field theory results}",
    eprint = "2008.01069",
    archivePrefix = "arXiv",
    primaryClass = "stat.ME",
    reportNumber = "FERMILAB-PUB-20-374-T",
    doi = "10.1103/PhysRevD.103.114502",
    journal = "Phys. Rev. D",
    volume = "103",
    pages = "114502",
    year = "2021"
}

@article{Pefkou:2021fni,
    author = "Pefkou, Dimitra A. and Hackett, Daniel C. and Shanahan, Phiala E.",
    title = "{Gluon gravitational structure of hadrons of different spin}",
    eprint = "2107.10368",
    archivePrefix = "arXiv",
    primaryClass = "hep-lat",
    reportNumber = "MIT-CTP/5318",
    doi = "10.1103/PhysRevD.105.054509",
    journal = "Phys. Rev. D",
    volume = "105",
    number = "5",
    pages = "054509",
    year = "2022"
}

@article{Brayshaw:1968yia,
    author = "Brayshaw, D. D.",
    title = "{Off- and on-shell analyticity of three-particle scattering amplitudes}",
    doi = "10.1103/PhysRev.176.1855",
    journal = "Phys. Rev.",
    volume = "176",
    pages = "1855--1870",
    year = "1968"
}

@article{Du:2021zzh,
    author = "Du, Meng-Lin and Baru, Vadim and Dong, Xiang-Kun and Filin, Arseniy and Guo, Feng-Kun and Hanhart, Christoph and Nefediev, Alexey and Nieves, Juan and Wang, Qian",
    title = "{Coupled-channel approach to Tcc+ including three-body effects}",
    eprint = "2110.13765",
    archivePrefix = "arXiv",
    primaryClass = "hep-ph",
    doi = "10.1103/PhysRevD.105.014024",
    journal = "Phys. Rev. D",
    volume = "105",
    number = "1",
    pages = "014024",
    year = "2022"
}

@article{Abolnikov:2024key,
    author = "Abolnikov, Michael and Baru, Vadim and Epelbaum, Evgeny and Filin, Arseniy A. and Hanhart, Christoph and Meng, Lu",
    title = "{Internal structure of the Tcc(3875)+ from its light-quark mass dependence}",
    eprint = "2407.04649",
    archivePrefix = "arXiv",
    primaryClass = "hep-ph",
    doi = "10.1016/j.physletb.2024.139188",
    journal = "Phys. Lett. B",
    volume = "860",
    pages = "139188",
    year = "2025"
}

@article{Sun:2024wxz,
    author = "Sun, Zhi-Feng and Li, Ning and Liu, Xiang",
    title = "{Isospin violation effect and three-body decays of the Tcc+ state}",
    eprint = "2405.00525",
    archivePrefix = "arXiv",
    primaryClass = "hep-ph",
    doi = "10.1103/PhysRevD.110.094025",
    journal = "Phys. Rev. D",
    volume = "110",
    number = "9",
    pages = "094025",
    year = "2024"
}

@article{Wang:2023iaz,
    author = "Wang, Jun-Zhang and Lin, Zi-Yang and Zhu, Shi-Lin",
    title = "{Cut structures and an observable singularity in the three-body threshold dynamics: The Tcc+ case}",
    eprint = "2309.09861",
    archivePrefix = "arXiv",
    primaryClass = "hep-ph",
    doi = "10.1103/PhysRevD.109.L071505",
    journal = "Phys. Rev. D",
    volume = "109",
    number = "7",
    pages = "L071505",
    year = "2024"
}

@article{Meng:2022ozq,
    author = "Meng, Lu and Wang, Bo and Wang, Guang-Juan and Zhu, Shi-Lin",
    title = "{Chiral perturbation theory for heavy hadrons and chiral effective field theory for heavy hadronic molecules}",
    eprint = "2204.08716",
    archivePrefix = "arXiv",
    primaryClass = "hep-ph",
    doi = "10.1016/j.physrep.2023.04.003",
    journal = "Phys. Rept.",
    volume = "1019",
    pages = "1--149",
    year = "2023"
}

@article{BaryonScattering:2025ziz,
    author = "Bulava, John and others",
    collaboration = "Baryon Scattering",
    title = "{Di-nucleons do not form bound states at heavy pion mass}",
    eprint = "2505.05547",
    archivePrefix = "arXiv",
    primaryClass = "hep-lat",
    reportNumber = "LLNL-JRNL-2005660",
    doi = "10.1103/d2hg-h6d4",
    journal = "Phys. Rev. C",
    volume = "113",
    number = "2",
    pages = "024002",
    year = "2026"
}

@article{Horz:2020zvv,
    author = {H{\"o}rz, Ben and others},
    title = "{Two-nucleon S-wave interactions at the $SU(3)$ flavor-symmetric point with $m_{ud}\simeq m_s^{\rm phys}$: A first lattice QCD calculation with the stochastic Laplacian Heaviside method}",
    eprint = "2009.11825",
    archivePrefix = "arXiv",
    primaryClass = "hep-lat",
    reportNumber = "LLNL-JRNL-813871, RIKEN-iTHEMS-Report-20, MITP/20-055",
    doi = "10.1103/PhysRevC.103.014003",
    journal = "Phys. Rev. C",
    volume = "103",
    number = "1",
    pages = "014003",
    year = "2021"
}

@article{Amarasinghe:2021lqa,
    author = "Amarasinghe, Saman and Baghdadi, Riyadh and Davoudi, Zohreh and Detmold, William and Illa, Marc and Parreno, Assumpta and Pochinsky, Andrew V. and Shanahan, Phiala E. and Wagman, Michael L.",
    title = "{Variational study of two-nucleon systems with lattice QCD}",
    eprint = "2108.10835",
    archivePrefix = "arXiv",
    primaryClass = "hep-lat",
    reportNumber = "FERMILAB-PUB-21-354-T, MIT-CTP/5320, UMD-PP-021-06",
    doi = "10.1103/PhysRevD.107.094508",
    journal = "Phys. Rev. D",
    volume = "107",
    number = "9",
    pages = "094508",
    year = "2023",
    note = "[Erratum: Phys.Rev.D 110, 119904 (2024)]"
}

@article{Detmold:2024iwz,
    author = "Detmold, William and Illa, Marc and Jay, William I. and Parre{\~n}o, Assumpta and Perry, Robert J. and Shanahan, Phiala E. and Wagman, Michael L.",
    collaboration = "NPLQCD",
    title = "{Constraints on the finite volume two-nucleon spectrum at m{\ensuremath{\pi}}{\ensuremath{\approx}}806{\,}{\,}MeV}",
    eprint = "2404.12039",
    archivePrefix = "arXiv",
    primaryClass = "hep-lat",
    reportNumber = "FERMILAB-PUB-24-0126-T, MIT-CTP/5700",
    doi = "10.1103/PhysRevD.111.114501",
    journal = "Phys. Rev. D",
    volume = "111",
    number = "11",
    pages = "114501",
    year = "2025"
}

@article{Yu:2025gzg,
    author = "Yu, Kang and Wang, Guang-Juan and Wu, Jia-Jun and Yang, Zhi",
    title = "{Finite volume Hamiltonian method for two-particle systems containing long-range potential on the lattice}",
    eprint = "2502.05789",
    archivePrefix = "arXiv",
    primaryClass = "hep-lat",
    doi = "10.1007/JHEP04(2025)108",
    journal = "JHEP",
    volume = "04",
    pages = "108",
    year = "2025"
}

@article{Wilson:2026bhu,
    author = "Wilson, David J. and Dudek, Jozef J. and Edwards, Robert G. and Thomas, Christopher E.",
    title = "{$D\bar{D}$ interactions are weak near threshold in QCD}",
    eprint = "2602.09862",
    archivePrefix = "arXiv",
    primaryClass = "hep-lat",
    reportNumber = "JLAB-THY-26-4597",
    month = "2",
    year = "2026"
}

@article{Stump:2025owq,
    author = "Stump, Andres and Green, Jeremy R.",
    title = "{Position-space sampling for local multiquark operators in lattice QCD using distillation and the importance of tetraquark operators for Tcc(3875)+}",
    eprint = "2510.26459",
    archivePrefix = "arXiv",
    primaryClass = "hep-lat",
    reportNumber = "HU-EP-25/35-RTG, DESY-25-140",
    doi = "10.1103/khhk-qw9n",
    journal = "Phys. Rev. D",
    volume = "113",
    number = "3",
    pages = "034513",
    year = "2026"
}

@article{Chen:2022vpo,
    author = "Chen, Siyang and Shi, Chunjiang and Chen, Ying and Gong, Ming and Liu, Zhaofeng and Sun, Wei and Zhang, Renqiang",
    title = "{Tcc+(3875) relevant DD{\textasteriskcentered} scattering from Nf = 2 lattice QCD}",
    eprint = "2206.06185",
    archivePrefix = "arXiv",
    primaryClass = "hep-lat",
    doi = "10.1016/j.physletb.2022.137391",
    journal = "Phys. Lett. B",
    volume = "833",
    pages = "137391",
    year = "2022"
}

@article{Briceno:2019nns,
    author = "Brice{\~n}o, Ra{\'u}l A. and Hansen, Maxwell T. and Jackura, Andrew W.",
    title = "{Consistency checks for two-body finite-volume matrix elements: I. Conserved currents and bound states}",
    eprint = "1909.10357",
    archivePrefix = "arXiv",
    primaryClass = "hep-lat",
    reportNumber = "JLAB-THY-19-3040, CERN-TH-2019-149",
    doi = "10.1103/PhysRevD.100.114505",
    journal = "Phys. Rev. D",
    volume = "100",
    number = "11",
    pages = "114505",
    year = "2019"
}

@article{Briceno:2015tza,
    author = "Brice{\~n}o, Ra{\'u}l A. and Hansen, Maxwell T.",
    title = "{Relativistic, model-independent, multichannel $2\to 2$ transition amplitudes in a finite volume}",
    eprint = "1509.08507",
    archivePrefix = "arXiv",
    primaryClass = "hep-lat",
    reportNumber = "JLAB-THY-15-2140",
    doi = "10.1103/PhysRevD.94.013008",
    journal = "Phys. Rev. D",
    volume = "94",
    number = "1",
    pages = "013008",
    year = "2016"
}

@article{Baroni:2018iau,
    author = "Baroni, Alessandro and Brice{\~n}o, Ra{\'u}l A. and Hansen, Maxwell T. and Ortega-Gama, Felipe G.",
    title = "{Form factors of two-hadron states from a covariant finite-volume formalism}",
    eprint = "1812.10504",
    archivePrefix = "arXiv",
    primaryClass = "hep-lat",
    reportNumber = "JLAB-THY-18-2878, CERN-TH-2018-263",
    doi = "10.1103/PhysRevD.100.034511",
    journal = "Phys. Rev. D",
    volume = "100",
    number = "3",
    pages = "034511",
    year = "2019"
}

@article{Edwards:2008ja,
    author = "Edwards, Robert G. and Joo, Balint and Lin, Huey-Wen",
    title = "{Tuning for Three-flavors of Anisotropic Clover Fermions with Stout-link Smearing}",
    eprint = "0803.3960",
    archivePrefix = "arXiv",
    primaryClass = "hep-lat",
    reportNumber = "JLAB-THY-08-806",
    doi = "10.1103/PhysRevD.78.054501",
    journal = "Phys. Rev. D",
    volume = "78",
    pages = "054501",
    year = "2008"
}

@article{HadronSpectrum:2008xlg,
    author = "Lin, Huey-Wen and others",
    collaboration = "Hadron Spectrum",
    title = "{First results from 2+1 dynamical quark flavors on an anisotropic lattice: Light-hadron spectroscopy and setting the strange-quark mass}",
    eprint = "0810.3588",
    archivePrefix = "arXiv",
    primaryClass = "hep-lat",
    reportNumber = "JLAB-THY-08-896, TCDMATH-08-13",
    doi = "10.1103/PhysRevD.79.034502",
    journal = "Phys. Rev. D",
    volume = "79",
    pages = "034502",
    year = "2009"
}

@article{Edwards:2012fx,
    author = "Edwards, Robert G. and Mathur, Nilmani and Richards, David G. and Wallace, Stephen J.",
    collaboration = "Hadron Spectrum",
    title = "{Flavor structure of the excited baryon spectra from lattice QCD}",
    eprint = "1212.5236",
    archivePrefix = "arXiv",
    primaryClass = "hep-ph",
    reportNumber = "JLAB-THY-12-1680, TIFR-TH-12-50",
    doi = "10.1103/PhysRevD.87.054506",
    journal = "Phys. Rev. D",
    volume = "87",
    number = "5",
    pages = "054506",
    year = "2013"
}

@article{HadronSpectrum:2012gic,
    author = "Liu, Liuming and Moir, Graham and Peardon, Michael and Ryan, Sinead M. and Thomas, Christopher E. and Vilaseca, Pol and Dudek, Jozef J. and Edwards, Robert G. and Joo, Balint and Richards, David G.",
    collaboration = "Hadron Spectrum",
    title = "{Excited and exotic charmonium spectroscopy from lattice QCD}",
    eprint = "1204.5425",
    archivePrefix = "arXiv",
    primaryClass = "hep-ph",
    reportNumber = "TCDMATH-12-04, JLAB-THY-12-1510",
    doi = "10.1007/JHEP07(2012)126",
    journal = "JHEP",
    volume = "07",
    pages = "126",
    year = "2012"
}

@article{HadronSpectrum:2009krc,
    author = "Peardon, Michael and Bulava, John and Foley, Justin and Morningstar, Colin and Dudek, Jozef and Edwards, Robert G. and Joo, Balint and Lin, Huey-Wen and Richards, David G. and Juge, Keisuke Jimmy",
    collaboration = "Hadron Spectrum",
    title = "{A Novel quark-field creation operator construction for hadronic physics in lattice QCD}",
    eprint = "0905.2160",
    archivePrefix = "arXiv",
    primaryClass = "hep-lat",
    reportNumber = "JLAB-THY-09-985",
    doi = "10.1103/PhysRevD.80.054506",
    journal = "Phys. Rev. D",
    volume = "80",
    pages = "054506",
    year = "2009"
}

@article{Michael:1985ne,
    author = "Michael, Christopher",
    title = "{Adjoint Sources in Lattice Gauge Theory}",
    reportNumber = "LTH 125",
    doi = "10.1016/0550-3213(85)90297-4",
    journal = "Nucl. Phys. B",
    volume = "259",
    pages = "58--76",
    year = "1985"
}

@article{Luscher:1990ck,
    author = "Luscher, Martin and Wolff, Ulli",
    title = "{How to Calculate the Elastic Scattering Matrix in Two-dimensional Quantum Field Theories by Numerical Simulation}",
    reportNumber = "DESY-90-010",
    doi = "10.1016/0550-3213(90)90540-T",
    journal = "Nucl. Phys. B",
    volume = "339",
    pages = "222--252",
    year = "1990"
}

@article{Blossier:2009kd,
    author = "Blossier, Benoit and Della Morte, Michele and von Hippel, Georg and Mendes, Tereza and Sommer, Rainer",
    title = "{On the generalized eigenvalue method for energies and matrix elements in lattice field theory}",
    eprint = "0902.1265",
    archivePrefix = "arXiv",
    primaryClass = "hep-lat",
    reportNumber = "DESY-09-014, SFB-CPP-09-10, MKPH-T-09-01, LPT-ORSAY-09-05",
    doi = "10.1088/1126-6708/2009/04/094",
    journal = "JHEP",
    volume = "04",
    pages = "094",
    year = "2009"
}

@article{Dudek:2010wm,
    author = "Dudek, Jozef J. and Edwards, Robert G. and Peardon, Michael J. and Richards, David G. and Thomas, Christopher E.",
    title = "{Toward the excited meson spectrum of dynamical QCD}",
    eprint = "1004.4930",
    archivePrefix = "arXiv",
    primaryClass = "hep-ph",
    reportNumber = "JLAB-THY-10-1171",
    doi = "10.1103/PhysRevD.82.034508",
    journal = "Phys. Rev. D",
    volume = "82",
    pages = "034508",
    year = "2010"
}

@article{Moscoso:2026wmz,
    author = "Moscoso, Joseph and Ortega-Gama, Felipe G. and Brice{\~n}o, Ra{\'u}l A. and Jackura, Andrew W. and Kacir, Charles and Nicholson, Amy N.",
    title = "{Resolving the structure of bound states using lattice quantum field theories}",
    eprint = "2602.20373",
    archivePrefix = "arXiv",
    primaryClass = "hep-lat",
    month = "2",
    year = "2026"
}

@article{Briceno:2025yuq,
    author = "Brice{\~n}o, Ra{\'u}l A. and Hansen, Maxwell T. and Jackura, Andrew W. and Edwards, Robert G. and Thomas, Christopher E.",
    title = "{Isotensor $πππ$ scattering with a $ρ$ resonant subsystem from QCD}",
    eprint = "2510.24894",
    archivePrefix = "arXiv",
    primaryClass = "hep-lat",
    reportNumber = "JLAB-THY-25-4592",
    month = "10",
    year = "2025"
}

@article{Dawid:2023kxu,
    author = "Dawid, Sebastian M. and Islam, Md Habib E. and Briceno, Raul A. and Jackura, Andrew W.",
    title = "{Evolution of Efimov states}",
    eprint = "2309.01732",
    archivePrefix = "arXiv",
    primaryClass = "nucl-th",
    doi = "10.1103/PhysRevA.109.043325",
    journal = "Phys. Rev. A",
    volume = "109",
    number = "4",
    pages = "043325",
    year = "2024"
}

@article{Raposo:2025dkb,
    author = "Raposo, Andr{\'e} Bai{\~a}o and Brice{\~n}o, Ra{\'u}l A. and Hansen, Maxwell T. and Jackura, Andrew W.",
    title = "{Extracting scattering amplitudes for arbitrary two-particle systems with one-particle left-hand cuts via lattice QCD}",
    eprint = "2502.19375",
    archivePrefix = "arXiv",
    primaryClass = "hep-lat",
    doi = "10.1007/JHEP06(2025)186",
    journal = "JHEP",
    volume = "06",
    pages = "186",
    year = "2025"
}

@article{LHCb:2021auc,
	title        = {{Study of the doubly charmed tetraquark $T_{cc}^{+}$}},
	author       = {Aaij, Roel and others},
	year         = 2022,
	journal      = {Nature Commun.},
	volume       = 13,
	number       = 1,
	pages        = 3351,
	doi          = {10.1038/s41467-022-30206-w},
	collaboration = {LHCb},
	eprint       = {2109.01056},
	archiveprefix = {arXiv},
	primaryclass = {hep-ex},
	reportnumber = {CERN-EP-2021-169, LHCb-PAPER-2021-032}
}

@article{LHCb:2021vvq,
	title        = {{Observation of an exotic narrow doubly charmed tetraquark}},
	author       = {Aaij, Roel and others},
	year         = 2022,
	journal      = {Nature Phys.},
	volume       = 18,
	number       = 7,
	pages        = {751--754},
	doi          = {10.1038/s41567-022-01614-y},
	collaboration = {LHCb},
	eprint       = {2109.01038},
	archiveprefix = {arXiv},
	primaryclass = {hep-ex},
	reportnumber = {CERN-EP-2021-165, LHCb-PAPER-2021-031}
}

@article{Hansen:2014eka,
    author = "Hansen, Maxwell T. and Sharpe, Stephen R.",
    title = "{Relativistic, model-independent, three-particle quantization condition}",
    eprint = "1408.5933",
    archivePrefix = "arXiv",
    primaryClass = "hep-lat",
    doi = "10.1103/PhysRevD.90.116003",
    journal = "Phys. Rev. D",
    volume = "90",
    number = "11",
    pages = "116003",
    year = "2014"
}

@article{Jackura:2020bsk,
    author = "Jackura, Andrew W. and Brice\~no, Ra\'ul A. and Dawid, Sebastian M. and Islam, Md Habib E. and McCarty, Connor",
    title = "{Solving relativistic three-body integral equations in the presence of bound states}",
    eprint = "2010.09820",
    archivePrefix = "arXiv",
    primaryClass = "hep-lat",
    reportNumber = "JLAB-THY-20-3272",
    doi = "10.1103/PhysRevD.104.014507",
    journal = "Phys. Rev. D",
    volume = "104",
    number = "1",
    pages = "014507",
    year = "2021"
}

@article{Dawid:2023jrj,
    author = "Dawid, Sebastian M. and Islam, Md Habib E. and Brice\~no, Ra\'ul A.",
    title = "{Analytic continuation of the relativistic three-particle scattering amplitudes}",
    eprint = "2303.04394",
    archivePrefix = "arXiv",
    primaryClass = "nucl-th",
    doi = "10.1103/PhysRevD.108.034016",
    journal = "Phys. Rev. D",
    volume = "108",
    number = "3",
    pages = "034016",
    year = "2023"
}

@article{Jackura:2023qtp,
    author = "Jackura, Andrew W. and Brice\~no, Ra\'ul A.",
    title = "{Partial-wave projection of the one-particle exchange in three-body scattering amplitudes}",
    eprint = "2312.00625",
    archivePrefix = "arXiv",
    primaryClass = "hep-ph",
    doi = "10.1103/PhysRevD.109.096030",
    journal = "Phys. Rev. D",
    volume = "109",
    number = "9",
    pages = "096030",
    year = "2024"
}

@article{Briceno:2024ehy,
    author = "Brice{\~n}o, Ra{\'u}l A. and Costa, Caroline S. R. and Jackura, Andrew W.",
    title = "{Partial-wave projection of relativistic three-body amplitudes}",
    eprint = "2409.15577",
    archivePrefix = "arXiv",
    primaryClass = "hep-ph",
    reportNumber = "JLAB-THY-24-4199",
    doi = "10.1103/PhysRevD.111.036029",
    journal = "Phys. Rev. D",
    volume = "111",
    number = "3",
    pages = "036029",
    year = "2025"
}

@article{Briceno:2020vgp,
    author = "Brice\~no, Ra\'ul A. and Jackura, Andrew W. and Ortega-Gama, Felipe G. and Sherman, Keegan H.",
    title = "{On-shell representations of two-body transition amplitudes: Single external current}",
    eprint = "2012.13338",
    archivePrefix = "arXiv",
    primaryClass = "hep-lat",
    reportNumber = "JLAB-THY-20-3298",
    doi = "10.1103/PhysRevD.103.114512",
    journal = "Phys. Rev. D",
    volume = "103",
    number = "11",
    pages = "114512",
    year = "2021"
}

@article{Raposo:2023oru,
    author = "Raposo, Andr\'e {Bai\~ao} and Hansen, Maxwell T.",
    title = "{Finite-volume scattering on the left-hand cut}",
    eprint = "2311.18793",
    archivePrefix = "arXiv",
    primaryClass = "hep-lat",
    doi = "10.1007/JHEP08(2024)075",
    journal = "JHEP",
    volume = "08",
    pages = "075",
    year = "2024"
}

@article{Christ:2005gi,
    author = "Christ, Norman H. and Kim, Changhoan and Yamazaki, Takeshi",
    title = "{Finite volume corrections to the two-particle decay of states with non-zero momentum}",
    eprint = "hep-lat/0507009",
    archivePrefix = "arXiv",
    reportNumber = "RBRC-530, SHEP-0520, CU-TP-1131",
    doi = "10.1103/PhysRevD.72.114506",
    journal = "Phys. Rev. D",
    volume = "72",
    pages = "114506",
    year = "2005"
}

@article{Kim:2005gf,
    author = "Kim, C. h. and Sachrajda, C. T. and Sharpe, Stephen R.",
    title = "{Finite-volume effects for two-hadron states in moving frames}",
    eprint = "hep-lat/0507006",
    archivePrefix = "arXiv",
    reportNumber = "UW-PT-05-16, SHEP-0518",
    doi = "10.1016/j.nuclphysb.2005.08.029",
    journal = "Nucl. Phys. B",
    volume = "727",
    pages = "218--243",
    year = "2005"
}

@article{Briceno:2012yi,
    author = "Briceno, Raul A. and Davoudi, Zohreh",
    title = "{Moving multichannel systems in a finite volume with application to proton-proton fusion}",
    eprint = "1204.1110",
    archivePrefix = "arXiv",
    primaryClass = "hep-lat",
    reportNumber = "NT-UW-12-05, NT@UW-12-05",
    doi = "10.1103/PhysRevD.88.094507",
    journal = "Phys. Rev. D",
    volume = "88",
    number = "9",
    pages = "094507",
    year = "2013"
}

@article{Hansen:2012tf,
    author = "Hansen, Maxwell T. and Sharpe, Stephen R.",
    title = "{Multiple-channel generalization of Lellouch-Luscher formula}",
    eprint = "1204.0826",
    archivePrefix = "arXiv",
    primaryClass = "hep-lat",
    doi = "10.1103/PhysRevD.86.016007",
    journal = "Phys. Rev. D",
    volume = "86",
    pages = "016007",
    year = "2012"
}

@article{Briceno:2014oea,
    author = "Briceno, Raul A.",
    title = "{Two-particle multichannel systems in a finite volume with arbitrary spin}",
    eprint = "1401.3312",
    archivePrefix = "arXiv",
    primaryClass = "hep-lat",
    reportNumber = "JLAB-THY-14-1833",
    doi = "10.1103/PhysRevD.89.074507",
    journal = "Phys. Rev. D",
    volume = "89",
    number = "7",
    pages = "074507",
    year = "2014"
}

@article{Dawid:2024oey,
    author = "Dawid, Sebastian M. and Jackura, Andrew W. and Szczepaniak, Adam P.",
    title = "{Finite-volume quantization condition from the N/D representation}",
    eprint = "2411.15730",
    archivePrefix = "arXiv",
    primaryClass = "hep-lat",
    reportNumber = "JLAB-THY-24-4216",
    doi = "10.1016/j.physletb.2025.139442",
    journal = "Phys. Lett. B",
    volume = "864",
    pages = "139442",
    year = "2025"
}

@article{Hansen:2024ffk,
    author = "Hansen, Maxwell T. and Romero-L{\'o}pez, Fernando and Sharpe, Stephen R.",
    title = "{Incorporating DD{\ensuremath{\pi}} effects and left-hand cuts in lattice QCD studies of the T$_{cc}$(3875)$^{+}$}",
    eprint = "2401.06609",
    archivePrefix = "arXiv",
    primaryClass = "hep-lat",
    reportNumber = "MIT-CTP/5667",
    doi = "10.1007/JHEP06(2024)051",
    journal = "JHEP",
    volume = "06",
    pages = "051",
    year = "2024"
}

@article{Dawid:2025wsn,
    author = "Dawid, Sebastian M. and Romero-L{\'o}pez, Fernando and Sharpe, Stephen R.",
    title = "{Comparison of integral equations used to study $ {T}_{cc}^{+} $ for a stable D$^{*}$}",
    eprint = "2505.05466",
    archivePrefix = "arXiv",
    primaryClass = "nucl-th",
    doi = "10.1007/JHEP09(2025)058",
    journal = "JHEP",
    volume = "09",
    pages = "058",
    year = "2025"
}

@article{Jacob:1959at,
    author = "Jacob, M. and Wick, G. C.",
    title = "{On the General Theory of Collisions for Particles with Spin}",
    doi = "10.1006/aphy.2000.6022",
    journal = "Annals Phys.",
    volume = "7",
    pages = "404--428",
    year = "1959"
}

@book{Martin:1970hmp,
    author = "Martin, A. D. and Spearman, T. D.",
    title = "{Elementary Particle Theory}",
    isbn = "978-0-7204-0157-8",
    publisher = "North-Holland Publishing Co.",
    address = "Amsterdam",
    year = "1970"
}

@article{Woss:2020cmp,
    author = "Woss, Antoni J. and Wilson, David J. and Dudek, Jozef J.",
    collaboration = "Hadron Spectrum",
    title = {{Efficient solution of the multichannel L{\"u}scher determinant condition through eigenvalue decomposition}},
    eprint = "2001.08474",
    archivePrefix = "arXiv",
    primaryClass = "hep-lat",
    reportNumber = "JLAB-THY-20-3133",
    doi = "10.1103/PhysRevD.101.114505",
    journal = "Phys. Rev. D",
    volume = "101",
    number = "11",
    pages = "114505",
    year = "2020"
}

@article{Woss:2018irj,
    author = "Woss, Antoni and Thomas, Christopher E. and Dudek, Jozef J. and Edwards, Robert G. and Wilson, David J.",
    title = "{Dynamically-coupled partial-waves in $\rho\pi$ isospin-2 scattering from lattice QCD}",
    eprint = "1802.05580",
    archivePrefix = "arXiv",
    primaryClass = "hep-lat",
    reportNumber = "DAMTP-2018-6, JLAB-THY-18-2643",
    doi = "10.1007/JHEP07(2018)043",
    journal = "JHEP",
    volume = "07",
    pages = "043",
    year = "2018"
}

@article{Dudek:2012gj,
    author = "Dudek, Jozef J. and Edwards, Robert G. and Thomas, Christopher E.",
    title = "{S and D-wave phase shifts in isospin-2 pi pi scattering from lattice QCD}",
    eprint = "1203.6041",
    archivePrefix = "arXiv",
    primaryClass = "hep-ph",
    reportNumber = "JLAB-THY-12-1504, TCDMATH-12-03",
    doi = "10.1103/PhysRevD.86.034031",
    journal = "Phys. Rev. D",
    volume = "86",
    pages = "034031",
    year = "2012"
}

@article{Wilson:2014cna,
    author = "Wilson, David J. and Dudek, Jozef J. and Edwards, Robert G. and Thomas, Christopher E.",
    title = "{Resonances in coupled $\pi K, \eta K$ scattering from lattice QCD}",
    eprint = "1411.2004",
    archivePrefix = "arXiv",
    primaryClass = "hep-ph",
    reportNumber = "JLAB-THY-14-1982, DAMTP-2014-81, JLAB-THY-14-1892",
    doi = "10.1103/PhysRevD.91.054008",
    journal = "Phys. Rev. D",
    volume = "91",
    number = "5",
    pages = "054008",
    year = "2015"
}

@article{PhysRevLett.129.252001,
  title = {Axial-Vector ${D}_{1}$ Hadrons in ${D}^{*}\ensuremath{\pi}$ Scattering from QCD},
  author = {Lang, Nicolas and Wilson, David J.},
  collaboration = {for the Hadron Spectrum Collaboration},
  journal = {Phys. Rev. Lett.},
  volume = {129},
  issue = {25},
  pages = {252001},
  numpages = {7},
  year = {2022},
  month = {Dec},
  publisher = {American Physical Society},
  doi = {10.1103/PhysRevLett.129.252001},
  url = {https://link.aps.org/doi/10.1103/PhysRevLett.129.252001}
}

@article{Lang:2025pjq,
    author = "Lang, Nicolas and Wilson, David J.",
    collaboration = "Hadron Spectrum",
    title = "{D$_{1}$ and D$_{2}$ resonances in coupled-channel scattering amplitudes from lattice QCD}",
    eprint = "2502.04232",
    archivePrefix = "arXiv",
    primaryClass = "hep-lat",
    doi = "10.1007/JHEP07(2025)060",
    journal = "JHEP",
    volume = "07",
    pages = "060",
    year = "2025"
}

@article{Becirevic:2012pf,
    author = "Becirevic, Damir and Sanfilippo, Francesco",
    title = "{Theoretical estimate of the $D^* \to D\pi$ decay rate}",
    eprint = "1210.5410",
    archivePrefix = "arXiv",
    primaryClass = "hep-lat",
    reportNumber = "LPT-ORSAY-12-105",
    doi = "10.1016/j.physletb.2013.03.004",
    journal = "Phys. Lett. B",
    volume = "721",
    pages = "94--100",
    year = "2013"
}

@article{Green:2021qol,
	title        = {{Weakly bound $H$ dibaryon from SU(3)-flavor-symmetric QCD}},
	author       = {Green, Jeremy R. and Hanlon, Andrew D. and Junnarkar, Parikshit M. and Wittig, Hartmut},
	year         = 2021,
	journal      = {Phys. Rev. Lett.},
	volume       = 127,
	number       = 24,
	pages        = 242003,
	doi          = {10.1103/PhysRevLett.127.242003},
	eprint       = {2103.01054},
	archiveprefix = {arXiv},
	primaryclass = {hep-lat},
	reportnumber = {MITP-21-009, CERN-TH-2021-024}
}

@article{Rodas:2026zmm,
    author = "Rodas, Arkaitz and Qiu, Lin and Fern{\'a}ndez-Ram{\'\i}rez, C{\'e}sar and Mathieu, Vincent and Monta{\~n}a, Gl{\`o}ria and Pilloni, Alessandro and Szczepaniak, Adam P.",
    title = "{Finite-volume analysis of the $H$-dibaryon including left-hand-cut effects}",
    eprint = "2605.22957",
    archivePrefix = "arXiv",
    primaryClass = "hep-lat",
    reportNumber = "JLAB-THY-26-4756",
    month = "5",
    year = "2026"
}

@article{Whyte:2024ihh,
    author = "Whyte, Travis and Wilson, David J. and Thomas, Christopher E.",
    collaboration = "Hadron Spectrum",
    title = "{Near-threshold states in coupled DD*-D*D* scattering from lattice QCD}",
    eprint = "2405.15741",
    archivePrefix = "arXiv",
    primaryClass = "hep-lat",
    doi = "10.1103/PhysRevD.111.034511",
    journal = "Phys. Rev. D",
    volume = "111",
    number = "3",
    pages = "034511",
    year = "2025"
}

@article{Padmanath:2022cvl,
    author = "Padmanath, M. and Prelovsek, S.",
    title = "{Signature of a Doubly Charm Tetraquark Pole in DD* Scattering on the Lattice}",
    eprint = "2202.10110",
    archivePrefix = "arXiv",
    primaryClass = "hep-lat",
    reportNumber = "MITP/22-018",
    doi = "10.1103/PhysRevLett.129.032002",
    journal = "Phys. Rev. Lett.",
    volume = "129",
    number = "3",
    pages = "032002",
    year = "2022"
}

@article{Collins:2024sfi,
    author = "Collins, Sara and Nefediev, Alexey and Padmanath, M. and Prelovsek, Sasa",
    title = "{Toward the quark mass dependence of Tcc+ from lattice QCD}",
    eprint = "2402.14715",
    archivePrefix = "arXiv",
    primaryClass = "hep-lat",
    reportNumber = "IMSc/24/01",
    doi = "10.1103/PhysRevD.109.094509",
    journal = "Phys. Rev. D",
    volume = "109",
    number = "9",
    pages = "094509",
    year = "2024"
}

@inproceedings{Alharazin:2026lno,
    author = "Alharazin, Herzallah and Raposo, Andr{\'e} Bai{\~a}o and Bulava, John and Dawid, Sebastian and Green, Jeremy R. and Morningstar, Colin and Romero-L{\'o}pez, Fernando and Salg, Miguel and Sharpe, Stephen R. and Stump, Andres",
    title = "{Three-body study of the $T_{cc}(3875)^+$ from lattice QCD}",
    booktitle = "{42th International Symposium on Lattice Field Theory}",
    eprint = "2602.17204",
    archivePrefix = "arXiv",
    primaryClass = "hep-lat",
    reportNumber = "DESY-26-021, HU-EP-26/10-RTG",
    month = "2",
    year = "2026"
}

@article{Lyu:2023xro,
    author = "Lyu, Yan and Aoki, Sinya and Doi, Takumi and Hatsuda, Tetsuo and Ikeda, Yoichi and Meng, Jie",
    title = "{Doubly Charmed Tetraquark Tcc+ from Lattice QCD near Physical Point}",
    eprint = "2302.04505",
    archivePrefix = "arXiv",
    primaryClass = "hep-lat",
    reportNumber = "RIKEN-iTHEMS-Report-23, YITP-23-14",
    doi = "10.1103/PhysRevLett.131.161901",
    journal = "Phys. Rev. Lett.",
    volume = "131",
    number = "16",
    pages = "161901",
    year = "2023"
}

@article{Prelovsek:2025vbr,
    author = "Prelovsek, Sasa and Ortiz-Pacheco, Emmanuel and Collins, Sara and Leskovec, Luka and Padmanath, M. and Vujmilovic, Ivan",
    title = "{Doubly heavy tetraquarks from lattice QCD: Incorporating diquark-antidiquark operators and the left-hand cut}",
    eprint = "2504.03473",
    archivePrefix = "arXiv",
    primaryClass = "hep-lat",
    doi = "10.1103/rlgp-c9tb",
    journal = "Phys. Rev. D",
    volume = "112",
    number = "1",
    pages = "014507",
    year = "2025"
}

@article{Du:2023hlu,
    author = "Du, Meng-Lin and Filin, Arseniy and Baru, Vadim and Dong, Xiang-Kun and Epelbaum, Evgeny and Guo, Feng-Kun and Hanhart, Christoph and Nefediev, Alexey and Nieves, Juan and Wang, Qian",
    title = "{Role of Left-Hand Cut Contributions on Pole Extractions from Lattice Data: Case Study for Tcc(3875)+}",
    eprint = "2303.09441",
    archivePrefix = "arXiv",
    primaryClass = "hep-ph",
    doi = "10.1103/PhysRevLett.131.131903",
    journal = "Phys. Rev. Lett.",
    volume = "131",
    number = "13",
    pages = "131903",
    year = "2023"
}

@article{Meng:2023bmz,
    author = "Meng, Lu and Baru, Vadim and Epelbaum, Evgeny and Filin, Arseniy A. and Gasparyan, Ashot M.",
    title = "{Solving the left-hand cut problem in lattice QCD: Tcc(3875)+ from finite volume energy levels}",
    eprint = "2312.01930",
    archivePrefix = "arXiv",
    primaryClass = "hep-lat",
    doi = "10.1103/PhysRevD.109.L071506",
    journal = "Phys. Rev. D",
    volume = "109",
    number = "7",
    pages = "L071506",
    year = "2024"
}

@article{Meng:2021uhz,
    author = "Meng, Lu and Epelbaum, E.",
    title = "{Two-particle scattering from finite-volume quantization conditions using the plane wave basis}",
    eprint = "2108.02709",
    archivePrefix = "arXiv",
    primaryClass = "hep-lat",
    doi = "10.1007/JHEP10(2021)051",
    journal = "JHEP",
    volume = "10",
    pages = "051",
    year = "2021"
}

@article{Bubna:2024izx,
    author = {Bubna, Rishabh and Hammer, Hans-Werner and M{\"u}ller, Fabian and Pang, Jin-Yi and Rusetsky, Akaki and Wu, Jia-Jun},
    title = {{L{\"u}scher equation with long-range forces}},
    eprint = "2402.12985",
    archivePrefix = "arXiv",
    primaryClass = "hep-lat",
    doi = "10.1007/JHEP05(2024)168",
    journal = "JHEP",
    volume = "05",
    pages = "168",
    year = "2024"
}

\clearpage
\onecolumngrid
\newpage


\appendix

\renewcommand{\appendixname}{}
\renewcommand{\thesection}{\Alph{section}}

\section*{Supplemental Material}

\section{Review of scattering formalism}
\label{sec:review}

In this section, we review the main ingredients needed from the formalism presented in Ref.~\cite{Raposo:2025dkb} for constraining the $DD^\star$ scattering amplitude from the finite-volume spectrum.
Reference~\cite{Raposo:2025dkb} showed that the scattering amplitude partial-wave-projected to a definite angular momentum $J$ and parity $P$ can be compactly written as Eq.~\ref{eq:main_amp}, which we rewrite here for convenience
\begin{align}
    \Mc^{J^P}=
    \Mc^{J^P}_\Ec + \Mc^{J^P}_{\Kc_0}  \,,
    \label{eq:app_amp}
\end{align}
where $\Mc^{J^P}_\Ec$ is determined by the partial-wave-projected one-pion exchange (OPE) $\Ec^{J^P}$ via a set of integral equations, and $\Mc^{J^P}_{\Kc_0}$ encodes the remaining short-range physics and can be written in terms of the partial-wave-projected matrices $\Kc_0^{J^P}$ and $\Mc^{J^P}_\Ec$.
In the limit where the OPE coupling vanishes, $\Mc^{J^P}_\Ec \to 0$ and $\Mc^{J^P}_{\Kc_0}$ reduces to the standard scattering amplitude.

The key point of Ref.~\cite{Raposo:2025dkb} is that $\Kc_0^{J^P}$ is the quantity that can be directly constrained from the finite-volume spectrum via the quantization condition,
\begin{align}
\det_{J m_J \ell} \big[\Kc_{0}^{-1} - i\rho + F + \Cc_L \big] 
=
\det_{J m_J \ell} \big[\Kc_{0}^{-1}  + \Fpv + \Cc_L \big]=0
\label{eq:QC_cc_spin} \,,
\end{align}
where $\rho$ is the two-body phase space, $F$ is the standard finite-volume function that appears in the literature~\cite{Kim:2005gf, Christ:2005gi, Hansen:2012tf, Briceno:2012yi, Briceno:2014oea}, and $\Cc_L$ is a new finite-volume function that encodes the effects of the OPE and vanishes when the OPE coupling is set to zero.
The determinant appearing in Eq.~\eqref{eq:QC_cc_spin} is over the total angular momentum $J$, its azimuthal component $m_J$, the orbital angular momentum $\ell$, and the intrinsic spin of the system $S$.
Because we are considering $DD^\star$ in the hadron-hadron elastic region, here $S$ will be fixed to $1$.
In the first equality, we have introduced a definition for $\Fpv = F - i\rho$, where the subscript alludes to the Principal value prescription.\footnote {It is important to note that the principal value prescription used in this definition is the standard one above threshold, but it differs below threshold. For a detailed discussion on this topic, we point the reader to Ref.~\cite{Hansen:2014eka}. }
This ensures that the quantization condition is given in terms of purely real functions.

Below, we give explicit expressions for all the building blocks of the finite-volume quantization condition and the infinite-volume amplitude.
By focusing on the applications for the $DD^\star$ system, we simplify some of the expressions appearing in Ref.~\cite{Raposo:2025dkb}.
Due to the reduction of rotational symmetry, the finite-volume functions will be written in a different basis than the infinite-volume amplitudes.

\subsection{Finite volume}
\label{sec:fvformalism}

In what follows, we will use ``$\star$" to denote variables defined in the center-of-momentum frame (CMF).
Let us first discuss the ingredients needed for the quantization condition, Eq.~\eqref{eq:QC_cc_spin}.
The standard finite-volume function $F$ for the $DD^\star$ system can be written as~\cite{Briceno:2014oea}
\begin{align}
F_{J'm_{J'}\ell';\, Jm_J\ell}   
&= \sum_{m,m',m_S}
\braket{\ell'm'1m_S|J'm_{J'}}
\braket{\ell m 1 m_S|J m_J}
F_{\ell'm';\, \ell m} \,,
\label{eq:Fjmls}
\end{align}
where 
\begin{align}
F_{\ell'm'; \ell m} (E, \mathbf{P}) 
& \equiv   \left[ \frac1{L^3} \sum_{\mathbf k} - \int \! 
\frac{d^3 \mathbf k}{(2\pi)^3} \right] \frac{\omega_{D}^\star}
{\omega_D(\mathbf k) }
\frac{ \Yc^T_{\ell' m'} (\mathbf k^\star, q^\star) \, 
\Yc_{\ell m}(\mathbf k^\star, q^\star)}
{2 E^\star \big[ q^{\star 2} - \mathbf k^{\star 2} + i\epsilon \big] } \,,
\label{eq:F_def}
\end{align}
$\omega_D(\mathbf k) = \sqrt{(\mathbf P-\mathbf k)^2+m_D^2}$, 
$\omega_{D}^\star = \sqrt{\mathbf k^{\star 2}+m_D^2}$, and $q^\star$ 
denotes the magnitude of the on-shell CMF momentum.
We have also introduced the spherical harmonics with barrier factors,
\begin{align}
\Yc_{\ell m}(\mathbf{k}^\star,q^\star) 
&\equiv \sqrt{4\pi} \left(\frac{\vert \mathbf{k}^\star \vert}{q^\star}
\right)^{\ell} Y_{\ell m}(\hat{\mathbf{k}}^\star) \,,
\label{eq:Yharm_def}
\\
\Yc_{\ell m}^T(\mathbf{k}^\star,q^\star) 
&\equiv \sqrt{4\pi} \left(\frac{\vert \mathbf{k}^\star \vert}{q^\star}
\right)^{\ell} Y^*_{\ell m}(\hat{\mathbf{k}}^\star) \,.
\label{eq:YharmT_def}
\end{align}

For $DD^\star$, the OPE receives contributions only from $u$-channel pion exchange, since the $t$-channel piece vanishes identically. 
Following the prescription of Ref.~\cite{Raposo:2025dkb}, we label all exchange contributions as $\Ec$. Furthermore, we symmetrize the OPE over the interchange of initial and final states, which ensures that all intermediate objects appearing in the formalism are symmetric.
With this, the exchange function for $DD^\star$ in the single-particle spin basis reads
\begin{align}
    \Ec_{\svec k' m'_{s_2}; \svec k m_{s_2}} (P)  &= 
    - \frac{g^2}{2} 
    \, \bbGamma_{m'_{s_2} m_{s_2}} (\svec k', \svec P -\svec k'; \svec k, \svec P -\svec k)
    \nonumber\\
    & \hspace{1cm}
    \times \Bigg( \frac{1}{u - m_\pi^2} \Bigg\vert_{\scriptsize \substack{k'^0=\omega_D(\svec k') \\ 
    k^0=E-\omega_{D^\star}(\svec P -\svec k)}}
    + \frac{1}{u - m_\pi^2} \Bigg \vert_{\scriptsize \substack{k'^0=E-\omega_{D^\star}(\svec P -\svec k') \\ 
    k^0=\omega_D(\svec k)}} \Bigg) \,,
    \label{eq:Ec_spin}
\end{align}
where $u=(P-k-k')^2$, $\bbGamma$ is a kinematic function carrying the Lorentz structure and $g$ is the dimensionless $D D^\star \pi$ coupling.
Here, $m_{s_2}$ ($m_{s_2}'$) is the azimuthal component of the $D^\star$ spin in the initial (final) state, along a fixed quantization axis.
The two terms correspond to the two kinematic prescriptions needed to avoid spurious singularities in the exchange function, as explained in Ref.~\cite{Raposo:2025dkb}, and their sum is symmetric under the interchange of initial and final states, $(\svec k, m_{s_2}) \leftrightarrow (\svec k', m'_{s_2})$.

We choose the function $\bbGamma$ to have its arguments completely on shell, i.e.
\begin{align}
    \bbGamma_{m'_{s_2} m_{s_2}} (\svec k_1', \svec k_2'; \svec k_1, \svec k_2) \equiv 
    k_1' \cdot \varepsilon(k_2, m_{s_2})
    \Big \vert_{\scriptsize \substack{
    k_1'^0=\omega_D(\svec k_1') \\
    k_2^0=\omega_{D^\star}(\svec k_2)
    }}
    \ k_1 \cdot \varepsilon^*(k_2', m'_{s_2}) 
    \Big \vert_{\scriptsize \substack{
    k_1^0=\omega_D(\svec k_1) \\ 
    k_2'^0=\omega_{D^\star}(\svec k_2')
    }}
     \,,
\end{align}
where $\varepsilon^\mu(p,m_s)$ are standard spin-$1$ polarization vectors for a massive particle in flight, quantized along a fixed axis.
In this way, the polarization vectors obey transversality $k^\mu \varepsilon_\mu(k,m_s) =0$, and orthogonality $\varepsilon^\mu(p, m_s) \varepsilon^*_\mu(p, m_s') = - \delta_{m_s m_s'}$.

Following the standard steps in refs.~\cite{Jacob:1959at,Martin:1970hmp}, we use the polarization vectors at rest and quantized along the $z$-axis,
\begin{equation}
    \varepsilon^\mu (\pm) = \frac1{\sqrt 2} (0, \mp 1, -i, 0) \,, 
    \qquad \varepsilon^\mu (0) = (0, 0, 0, 1) \,,
\end{equation}
to construct the polarization vectors in flight,
\begin{equation}
    \varepsilon^\mu(p,m_s) = \Lambda^\mu_\nu(p) \ \varepsilon^\nu 
    (m_s) \,,
\end{equation}
where $\Lambda(p)$ is an active Lorentz boost along $\svec{\hat p} = \svec p / |\svec p|$, which in Cartesian coordinates reads
\begin{equation}
    \Lambda^0_0 = \gamma \,, \quad
    \Lambda^0_i = \Lambda^i_0 = \gamma \beta (\svec{\hat p})_i \,,
    \quad
    \Lambda^i_j = \delta_{ij} + (\gamma - 1) (\svec{\hat p})_i 
    (\svec{\hat p})_j \,,
\end{equation}
with $\beta=|\svec p|/\omega$ and $\gamma = \omega/m$ for a particle of mass $m$ and energy $\omega = \sqrt{m^2 + |\svec p|^2}$.
We refer to the polarization vectors obtained in this way as being in the canonical basis.

Once we have the OPE, we will need to introduce a finite-volume two-particle propagator, which has the form
\begin{align}
(\Delta_{2,L})_{m_{s_2}'; m_{s_2}} (\mathbf k', \mathbf k)
&\equiv 
\frac1{L^3} \, \frac{\omega_{D}^\star}{\omega_D(\mathbf k) }
\frac{\delta_{m_{s_2}'m_{s_2}} \, \delta_{\mathbf k' \mathbf k} \, 
H (k^\star) }{2 E^\star \big[ q^{\star 2} - \mathbf k^{\star 2} 
 \big] } \,,
\label{eq:Delta2L_def_ms}
\end{align}
where $H(\mathbf{k}^\star)$ is a cutoff 
function that must be equal to $1$ at the on-shell point. In practice, 
we take the cutoff function to be \begin{align}
H(\mathbf{k}^\star) = 
\exp[-\alpha(k^{\star 2} - q^{\star 2})] \,,
\label{eq:Hfunc}
\end{align}
for some small volume-independent parameter $\alpha$.
This parameter plays a critical role in this analysis.
As discussed in Sec.~\ref{sec:specanalysis}, we fit the spectrum for a range of values of $\alpha$, and in Sec.~\ref{sec:alphavariation} we explain in some detail the choices of $\alpha$ made. 

Given $\Ec$ and $\Delta_{2,L}$, the finite-volume ladder sum can be solved in closed form,
\begin{equation}
\Mc_{\Ec,L} = \Ec\left[ 1 + \Delta_{2,L} \, \Ec\right]^{-1} \,,
\label{eq:MctL_sol_spin}
\end{equation}
where all objects are understood as matrices in the combined space of finite-volume momenta and $D^\star$ spin indices.
From this, the new finite-volume function entering the quantization condition is
\begin{align}
(i\Cc_L)_{\ell' m' m'_{s_2};\,\ell m m_{s_2}}
&\equiv 
\left( \Yc^T \, i\Delta_{2,L} \, i\Mc_{\Ec,L} \, i\Delta_{2,L} \, 
\Yc \right)_{\ell' m' m'_{s_2};\,\ell m m_{s_2}} \,,
\label{eq:CL_spin}
\end{align}
where the spherical harmonics act only on the orbital indices and the spin indices are carried through by $\Delta_{2,L}$ and $\Mc_{\Ec,L}$.
In order to implement Eq.~\eqref{eq:CL_spin}, one must understand the modified spherical harmonics as asymmetric tensors that allow one to go from momentum space to angular momentum indices.
For an explicit explanation of this compact matrix product notation, we point the reader to the original paper~\cite{Raposo:2025dkb}.

Just like the $F$ function, $\Cc_L$ can then be projected to a definite total angular 
momentum $J$ using standard Clebsch-Gordan coefficients,
\begin{align}
(\Cc_L)_{J'm_{J'}\ell';\, Jm_J\ell}  
&= \sum_{m,m',m_S,m_S'}
\braket{\ell'm'1m_S'|J'm_{J'}}
\braket{\ell m 1 m_S|J m_J}
(\Cc_L)_{\ell' m' m'_{S};\,\ell m m_{S}} \,.
\label{eq:CLjmls}
\end{align}
Note that in this work, $S'=S=1$ and so $m'_S = m'_{s_2}$ and $m_S = m_{s_2}$.

Having written $F$ and $\Cc_L$ in the ``$J m_J \ell S$'' basis in Eqs.~\eqref{eq:Fjmls} and~\eqref{eq:CLjmls}, standard techniques can be used to subduce those into lattice irreps~\cite{Dudek:2012gj}. This enables us to constrain $\Kc_0$ partial waves from the subduced version of the quantization condition Eq.~\eqref{eq:QC_cc_spin} in analogy to the usual two-particle formalism, see Sec.~\ref{sec:specanalysis}.


\subsection{Infinite volume}
\label{sec:amps}

Given that $\Kc_0^{J^P}$ can be obtained from the quantization condition Eq.~\eqref{eq:QC_cc_spin} for some total angular momentum and parity $J^P$, the physical scattering amplitude can then be reconstructed via Eq.~\eqref{eq:app_amp}. Let us now define the two terms appearing in that equation, starting with the second one,
\begin{align}  
   \Mc^{J^P}_{\Kc_0} \equiv
    \left(1+i\Lc^{J^P}\right) \left[\left(\Kc_0^{J^P}\right)^{-1} + \Cc^{J^P}
    -i \rho
    \right]^{-1} 
    \left(1+i\Rc^{J^P}\right) \,,
\end{align}
where $\rho = q^\star / (8\pi \sqrt{s})$.
The rest of the quantities can be defined in terms of the infinite-volume ladder amplitude, $\Mc_{\Ec}^{J^P}$, which satisfies the integral equation
\begin{align}
\big( i \Mc_{\Ec}^{J^P} \big)_{\ell';\,\ell} (p',p)
&= i\Ec_{\ell';\,\ell}^{J^P} (p',p) \nonumber\\
&\hspace{0.5cm}+ \sum_{\ell''} \int_0^\infty \frac{k^2 \,\mathrm{d}k}{2\pi^2}\, 
i\Ec_{\ell';\,\ell'' }^{J^P}(p',k)
\,(i\Delta_{2,\infty})(k)
\,\big( i\Mc_\Ec^{J^P} \big)_{\ell'' ;\,\ell}(k,p) \,,
\label{eq:Mt_int_spin}
\end{align}
where $\Ec^{J^P}$ is the partial wave projected OPE, and
\begin{align}
\Delta_{2,\infty}( k)
&\equiv 
 \, 
\frac{ \, 
H (k^\star) }{2 \sqrt{s} \big[ q^{\star 2} -  k^2 
+ i\epsilon \big] } \,.
\label{eq:Delta2_infty}
\end{align}

The remaining quantities can be written in terms of $\Mc_{\Ec}^{J^P}$ as
\begin{align}
i\Rc^{J^P}_{\ell';\,\ell} (p)
&= \int_{0}^\infty \frac{k^2 \,\mathrm{d}k}{2\pi^2} 
\left[\frac{k}{q^\star} \right]^{\ell'} 
(i\Delta_{2,\infty})(k) \, 
\big( i\Mc_\Ec^{J^P} \big)_{\ell';\,\ell} (k, p) \,,
\label{eq:iR}
\\
i\Lc^{J^P}_{\ell';\,\ell} (p')
&= \int_{0}^\infty \frac{k^2 \,\mathrm{d}k}{2\pi^2} \,
\big( i\Mc_\Ec^{J^P} \big)_{\ell';\,\ell} (p',k) \,
(i\Delta_{2,\infty})(k)
\left[\frac{k}{q^\star} \right]^{\ell} \,, 
\label{eq:iL}
\\
i\Cc^{J^P}_{\ell';\,\ell}
&= \int_{0}^\infty \frac{k'^2 \,\mathrm{d}k'}{2\pi^2}
\int_{0}^\infty \frac{k^2 \,\mathrm{d}k}{2\pi^2}
\left[\frac{k'}{q^\star} \right]^{\ell'}
(i\Delta_{2,\infty})(k')
\, \big( i\Mc_\Ec^{J^P} \big)_{\ell';\,\ell}(k', k) \,
(i\Delta_{2,\infty})(k)
\left[\frac{k}{q^\star} \right]^{\ell} \,.
\label{eq:C}
\end{align}

The only quantity that needs to be specified as an input to these equations is the partial-wave-projected OPE, $\Ec^{J^P}$, which we discuss in the next subsection.

\begin{figure}[htbp]
    \centering
    \includegraphics[width=.85\textwidth]{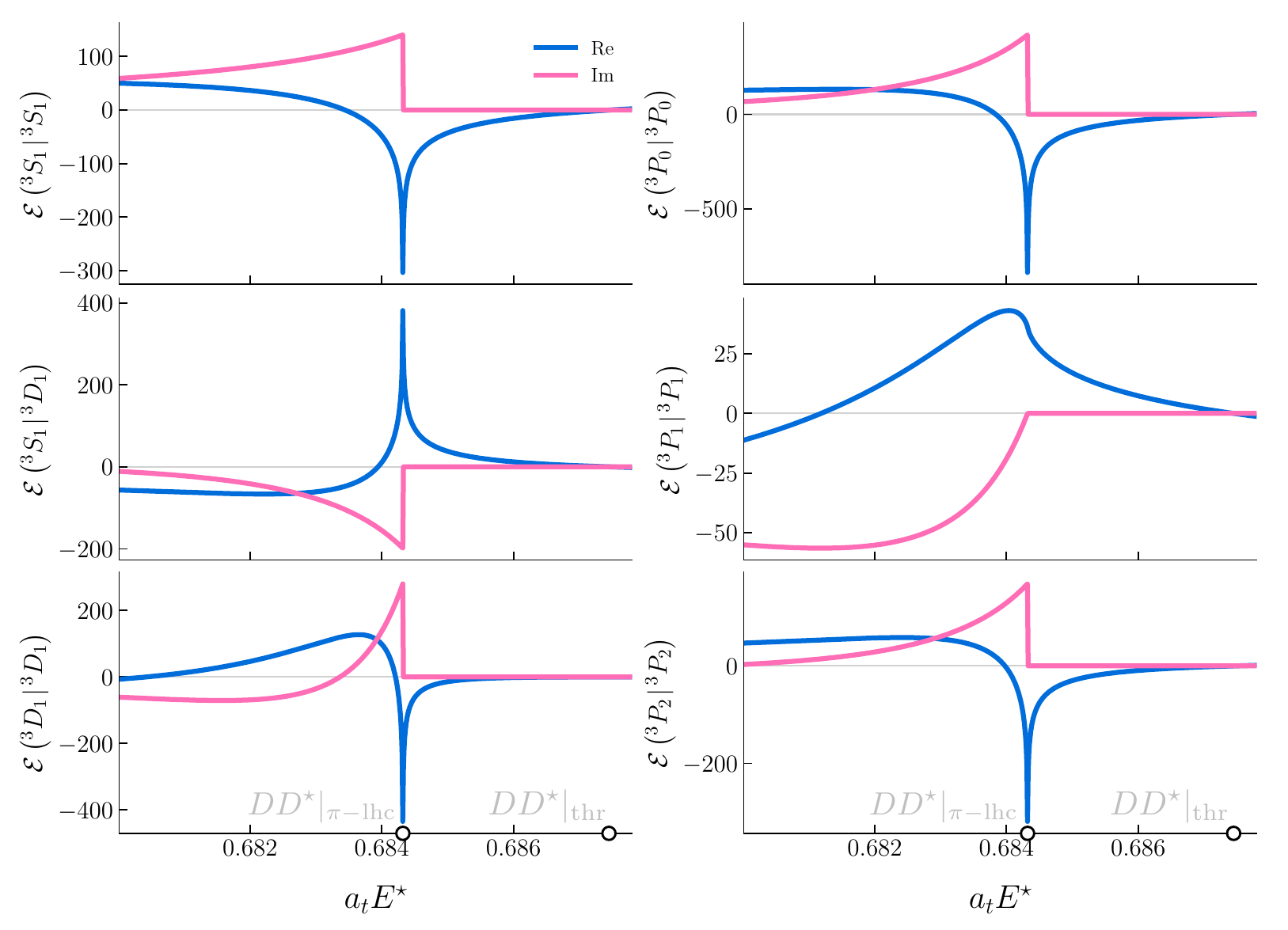}
     \caption{
    Example of partial-wave-projected OPE contribution used in the integral equation for $g=12$.
     }
    \label{fig:OPE}
\end{figure}

\subsection{Partial-wave-projected OPE for $J^P = 1^+$}
\label{sec:OPE_1plus}

Following Ref.~\cite{Raposo:2025dkb}, we write the integral equations in the partial-wave-projected basis in the CMF.
There, it was shown that the driving term in the integral equations takes the form
\begin{align}
    \Ec^{J^P} &= \frac{1}{2}\left[ \Cc_1^{J^P}
    +\Cc_2^{J^P}Q_0(\zeta + i\epsilon) + 
    (\mathbf{p}_i^\star \leftrightarrow \mathbf{p}_f^\star)\right] \,,
    \label{eq:Ec_pw}
\end{align}
where the second term between brackets is obtained from the first by swapping the initial and final CMF momenta, $\mathbf{p}_i^\star \leftrightarrow \mathbf{p}_f^\star$.
Below, we give explicit expressions for the first term only, as the second follows straightforwardly from it. Here, $Q_0$ is the Legendre function of the second kind, whose branch points at $\zeta = \pm 1$ generate the left-hand cut in the complex-$s$ plane, and
\begin{align}
\zeta &= -1-\frac{m_\pi^2 - u_0}{2|\textbf{p}_i^\star||\textbf{p}_f^\star|} \,,
\label{eq:zeta}
\\
u_0 &= m_D^2 + m_{D^*}^2 
- 2\,\omega_{D}(\textbf{p}_i^{\star 2})\,\omega_{D^*}(\textbf{p}_f^{\star 2}) 
+ 2\,|\textbf{p}_i^\star|\,|\textbf{p}_f^\star|\,,
\end{align}
where $\mathbf{p}_i^\star$ and $\mathbf{p}_f^\star$ are the initial and final relative momenta in the CMF, respectively, and $\Cc_1^{J^P}$ and $\Cc_2^{J^P}$ are kinematic coefficient matrices in the $\ell$ space, to be specified for each $J^P$.
In this work, we restrict our attention to $J^P=1^+$, where both ${}^3S_1$ and ${}^3D_1$ contribute.

To concisely write these coefficients, we first define a set of common building blocks.
To simplify the expressions, we introduce a labeling notation used in Ref.~\cite{Raposo:2025dkb}.
We label the incoming $D$ and $D^\star$ as particles 1 and 2, and the outgoing $D$ and $D^\star$ as particles 3 and 4, respectively.
Then, the kinematic quantities $\omega_j$ and $\beta_j$ refer to the energies and boost factors of these particles in the CMF, with
$|\svec{p}_1^\star| = 
|\svec{p}_2^\star| = |\svec{p}_i^\star|$ and $|\svec{p}_3^\star| = 
|\svec{p}_4^\star| = |\svec{p}_f^\star|$.
With this, we can define for the $j^{\rm th}$ particle its Lorentz factors $\gamma_j = \omega_j / m_j$ and boost factors $\beta_j = |\svec{p}_j|/\omega_j$.

Given these kinematic variables, we introduce four linear combinations,
\begin{align}
    f_1 &= \gamma_1 \left( \frac{\beta_1}{\beta_4} + \zeta \right) \,, \\
    f_3 &= \gamma_3 \left( \frac{\beta_3}{\beta_2} + \zeta \right) \,,
    \\
    f &= f_1 + f_3 - \zeta\,, \\
    f' &= \gamma_1\gamma_3 \left[ \zeta \left( \frac{\beta_1}{\beta_4} + \frac{\beta_3}{\beta_2} + \zeta \right) + \frac{\beta_1\beta_3}{\beta_4\beta_2} \right] \,.
\end{align}
With these, the coefficients $\Cc_1^{1^+}$ and $\Cc_2^{1^+}$ allow transitions between the ${}^3S_1$ and ${}^3D_1$ and can be written as
\begin{subequations}
\label{eq:all_amplitudes}
\begin{align}
    C_1^{1^+}({}^3S_1\to{}^3S_1) &= -\frac{g^2}{6}\left[ 
    \frac{2}{3}(1-\gamma_1)(1-\gamma_3) - \gamma_1\gamma_3 + 
    \zeta f - f' \right] \,, \\
    C_2^{1^+}({}^3S_1\to{}^3S_1) &= \frac{g^2}{6} \left[ 
    (\zeta^2 - 1)f - \zeta f' \right] \,, \\
    C_1^{1^+}({}^3S_1\to{}^3D_1) &= -\frac{g^2}{6\sqrt{2}}\left[ 
    \frac{2}{3}(1+2\gamma_3)(1-\gamma_1) + 2\gamma_1\gamma_3 + 
    \zeta(-2f_3 + f_1 - \zeta) + 2f' \right] \,, \\
    C_2^{1^+}({}^3S_1\to{}^3D_1) &= -\frac{g^2}{6\sqrt{2}}\left[ 
    (1-\zeta^2)(-2f_3 + f_1 - \zeta) - 2\zeta f' \right]  \,,\\
    C_1^{1^+}({}^3D_1\to{}^3D_1) &= -\frac{g^2}{12}\left[ \frac2{3}(1+2\gamma_1)(1+2\gamma_3) - 4\gamma_1\gamma_3 - \zeta (2 f+\zeta) - 4f' \right] \,, \\
    C_2^{1^+}({}^3D_1\to{}^3D_1) &= \frac{g^2}{12} \left[ (1-\zeta^2)(2 f+3\zeta) - 4\zeta f'\right] \,.
\end{align}
\end{subequations}
Note that the diagonal components are symmetric under the exchange of particle indices $1 \leftrightarrow 3$, while the off-diagonal transitions are related by a transpose symmetry, i.e.~the ${}^3D_1\to{}^3S_1$ coefficients are obtained from those for ${}^3S_1\to{}^3D_1$ by swapping particles $1 \leftrightarrow 3$.

\subsection{Partial-wave-projected OPE for $P$ waves}
\label{sec:OPE_pwaves}

Using the same notation and building blocks as above, we can compactly write the diagonal $P$-wave OPE coefficients for the ${}^3P_0$, ${}^3P_1$, and ${}^3P_2$ channels.
For these partial waves, the coefficients are
\begin{align}
\label{eq:ope_coeffs_3p0}
    C_1^{0^-}({}^3P_0\to{}^3P_0)
    &= \frac{g^2}{2}\gamma_1\gamma_3
    \left(
        \frac{\beta_1}{\beta_4}
        + \frac{\beta_3}{\beta_2}
        + \zeta
    \right) \,, \\
    C_2^{0^-}({}^3P_0\to{}^3P_0)
    &= -\frac{g^2}{2}\gamma_1\gamma_3
    \left(
        \frac{\beta_3}{\beta_2} + \zeta
    \right)
    \left(
        \frac{\beta_1}{\beta_4} + \zeta
    \right) \,, \\
    \label{eq:ope_coeffs_3p1}
    C_1^{1^-}({}^3P_1\to{}^3P_1)
    &= \frac{g^2}{4}\zeta \,, \\
    C_2^{1^-}({}^3P_1\to{}^3P_1)
    &= \frac{g^2}{4}(1-\zeta^2) \,, \\
    \label{eq:ope_coeffs_3p2}
    C_1^{2^-}({}^3P_2\to{}^3P_2)
    &= \frac{g^2}{20}
    \left[
        2(2-3\zeta^2)f
        +6\zeta f'
        -3\zeta
    \right] \,, \\
    C_2^{2^-}({}^3P_2\to{}^3P_2)
    &= \frac{g^2}{20}
    \left(
        3(\zeta^2-1)
        \left[1-2\zeta^2+2\zeta f_3\right)
        +2f_1
        \left(
            3\zeta(\zeta^2-1)
            +(1-3\zeta^2)f_3
        \right)
    \right] \,.
\end{align}
The expression for $C_2^{2^-}$ is not manifestly symmetric in $f_1$ and $f_3$ because it corresponds to the first term in Eq.~\eqref{eq:Ec_pw}; the full kernel has the required symmetry after adding the exchanged
contribution.

In Fig.~\ref{fig:OPE}, we give an example of the OPE evaluated for all partial waves considered here, assuming that the coupling is fixed to $g=12$.
For simplicity, we fix the initial and final state momenta to their on-shell values.
As one can see, these clearly exhibit the minimal criteria required, which include the explicit manifestation of the left-hand cut and their near-threshold behavior, which at most can scale as $\left(q^{\star}\right)^{\ell +\ell'}$. 
Given these and the determination of $\Kc_0$, we can proceed to solve the integral equation of Eq.~\eqref{eq:Mt_int_spin} using standard techniques laid out in the literature~\cite{Jackura:2020bsk, Dawid:2023jrj, Briceno:2024ehy}.
We next discuss our constraints on $\Kc_0$ from the finite-volume fits, before discussing the amplitude results in Sec.~\ref{sec:scatamps_poles}.
\\
  
\clearpage
\section{Spectrum analysis}
\label{sec:specanalysis}

We reanalyze the isoscalar $DD^\star$ spectrum computed in Ref.~\cite{Whyte:2024ihh} below the $D^\star D^\star$ threshold, corresponding to the $36$ finite-volume energy levels up to $a_t E^\star = 0.705$.
The levels are distributed across three spatial volumes $L/a_s = \{ 16,20,24 \}$ in the rest-frame irreps $T_1^+, A_1^-$ and $E^-$ labeled by parity, and in the moving-frame irreps $[001]A_2, [002]A_2, [011]A_2$ and $[111] A_2$ labeled by the total momentum $\svec{d} = [n_x n_y n_z]$ in units of $2\pi/L$.
As in Ref.~\cite{Whyte:2024ihh}, we adopt the same systematic error of $a_t \delta_{\mathrm syst} = 5 \times 10^{-4}$ in all levels to account for deviations from the continuum dispersion relation in the single-hadron sector.

We proceed to determine the energy dependence of $\Kc_0$ through appropriate parametrizations.
First, note that analyticity requires the components of $\Kc_0$ to vanish near threshold as $(\Kc_0^{J^P})_{\ell',\ell} \sim \left(q^{\star}\right)^{\ell +\ell'}$.
This motivates us to introduce a new function, $\Kpar_0$, that has barrier factors stripped out, 
\begin{align} 
    (\Kpar_0)_{\ell',\ell} 
    \equiv 
        \frac{1}{16 \pi} \frac{(\Kc_0)_{\ell',\ell}  }{\left(2q^{\star}\right)^{\ell +\ell'}} \,,
        \label{eq:Kc0bar}
\end{align}
following the conventions of Ref~\cite{Whyte:2024ihh}, where we have leave implicit the $J^P$ superscript and the constant total spin quantum number.
We then parametrize $\Kpar_0$ with polynomials in powers of Mandelstam $s$ around ${s_\mathrm{thr} = (m_D + m_{D^\star})^2}$,
\begin{equation}
    (\Kpar_0)_{ij}(s) = \sum_{n} \lhcgamma_{ij}^{(n)} (s - s_\mathrm{thr})^n \,,
    \label{eq:k_poly}
\end{equation}
where the compound indices $i$ and $j$ label all the quantum numbers of the final and initial states, respectively.
We also parametrize the inverse of $\Kpar_0$ as a polynomial,
\begin{equation}
    (\Kpar_0^{-1})_{ij}(s) = \sum_{n} \lhcinvkc_{ij}^{(n)} (s - s_\mathrm{thr})^n \,.
    \label{eq:kinv_poly}
\end{equation}
to reduce the bias coming from a single choice of parametrization form.
Note that in the limit $g\to0$, the Lüscher formalism is recovered and $\Kc_0$ approaches the physical K-matrix, and so $\lhcgamma \to \gamma$ and $\lhcinvkc \to c$, with $\gamma$ and $c$ as defined in Refs.~\cite{Wilson:2014cna,Whyte:2024ihh}.

We focus on parametrizations with $J^P = 1^+$, but due to possible partial-wave mixing in finite volume, we also examine contributions from $J^P = 0^-, 2^\pm, 3^+$ up to $D$-waves, which are the leading contributions to the irreps considered.
Table~\ref{tab:subduction} details the partial wave subductions to the irreps considered in this work.

\begin{table*}[!h]
\centering
{\small
\setlength{\tabcolsep}{5pt}
\renewcommand{\arraystretch}{1.75}

\newcommand{\inflatestrut}{\rule{0pt}{3.2ex}}

\begin{tabular}{|c|c|c|c|c|c|}
\hline
$T^+_1$ & $A^-_1$ & $E^-$ 
& $[00n]A_2$ & $[0nn]A_2$ & $[nnn]A_2$ \\
\hline\hline

 & $0^- \left( \SLJ{3}{P}{0} \right)$ & 
& \inflatestrut $0^- \left( \SLJ{3}{P}{0} \right)$ 
& \inflatestrut $0^- \left( \SLJ{3}{P}{0} \right)$ 
& \inflatestrut $0^- \left( \SLJ{3}{P}{0} \right)$ \\

$1^+ \left( \begin{matrix}\SLJ{3}{S}{1}\\ \SLJ{3}{D}{1}\end{matrix} \right)$ 
& & 
& \inflatestrut $1^+ \left( \begin{matrix}\SLJ{3}{S}{1}\\ \SLJ{3}{D}{1}\end{matrix} \right)$ 
& \inflatestrut $1^+ \left( \begin{matrix}\SLJ{3}{S}{1}\\ \SLJ{3}{D}{1}\end{matrix} \right)$ 
& \inflatestrut $1^+ \left( \begin{matrix}\SLJ{3}{S}{1}\\ \SLJ{3}{D}{1}\end{matrix} \right)$ \\

& & & & \inflatestrut $2^+ \left( \SLJ{3}{D}{2} \right)$
& \\

& & $2^- \left( \begin{matrix}\SLJ{3}{P}{2}\\ \textcolor{gray}{\SLJ{3}{F}{2}}\end{matrix} \right)$ 
& \inflatestrut $2^- \left( \begin{matrix}\SLJ{3}{P}{2}\\ \textcolor{gray}{\SLJ{3}{F}{2}}\end{matrix} \right)$ 
& \inflatestrut $2^- \left( \begin{matrix}\SLJ{3}{P}{2}\\ \textcolor{gray}{\SLJ{3}{F}{2}}\end{matrix} \right)_{[2]}$ 
& \inflatestrut $2^- \left( \begin{matrix}\SLJ{3}{P}{2}\\ \textcolor{gray}{\SLJ{3}{F}{2}}\end{matrix} \right)$ \\

$3^+ \left( \begin{matrix}\SLJ{3}{D}{3}\\ \textcolor{gray}{\SLJ{3}{G}{3}}\end{matrix} \right)$ 
& & 
& \inflatestrut $3^+ \left( \begin{matrix}\SLJ{3}{D}{3}\\ \textcolor{gray}{\SLJ{3}{G}{3}}\end{matrix} \right)$ 
& \inflatestrut $3^+ \left( \begin{matrix}\SLJ{3}{D}{3}\\ \textcolor{gray}{\SLJ{3}{G}{3}}\end{matrix} \right)_{[2]}$ 
& \inflatestrut $3^+ \left( \begin{matrix}\SLJ{3}{D}{3}\\ \textcolor{gray}{\SLJ{3}{G}{3}}\end{matrix} \right)_{[2]}$ \\

 & & 
& \inflatestrut 
& \inflatestrut $3^- \left( \textcolor{gray}{\SLJ{3}{F}{3}} \right)$ 
& \inflatestrut $3^- \left( \textcolor{gray}{\SLJ{3}{F}{3}} \right)$ \\
\hline
\end{tabular}
}
\caption{\label{tab:subduction}
Subduction table for a $DD^\star$ system in the rest and moving-frame irreps used in this work, up to $J=3$ and $\ell=4$.
The allowed partial-wave components and their respective $J^P$ are given using the notation $J^P(\SLJ{2S+1}{\ell}{J})~$, where $S=1$ for our case.
The subscript square brackets indicate the number of times a given partial wave appears subduced into the given irrep (number of embeddings).
The partial waves in gray are given for completeness but are not considered in the analysis.}
\end{table*}

\subsection{Reference parametrization for $g$ fixed to zero}
\label{sec:refluscher}

For comparison purposes, we give here a parametrization obtained at fixed $g=0$, i.e., using the Lüscher quantization condition. 
We will use the same form as that found in Ref.~\cite{Whyte:2024ihh} as a reasonable description of the $36$ finite-volume energies below the $D^\star D^\star$ threshold, assuming the absence of left-hand cut effects,\footnote{We use a simple two-body phase space instead of the Chew-Mandelstam prescription, the latter employed in the reference parametrization of Ref.~\cite{Whyte:2024ihh} for energies below $D^\star D^\star$. The following qualitative results are unchanged.} namely a polynomial parametrization given by
\begin{equation}
\renewcommand{\arraystretch}{1.5}
    \begin{tabular}{rll}
    $\gamma_{\SLJ{3}{D}{3}}^{(0)} =$ & $(-514 \pm 240 ) \cdot a_t^4$ & \multirow{7}{*}{ $\begin{bmatrix*}[r]   1.00 &   0.16 &  -0.26 &   0.23 &  -0.60 &   0.08 &  -0.05\\
    &  1.00 &   0.21 &   0.74 &   0.41 &   0.65 &  -0.57\\
    &&  1.00 &   0.08 &   0.08 &   0.19 &  -0.18\\
    &&&  1.00 &   0.48 &   0.63 &  -0.56\\
    &&&&  1.00 &   0.40 &  -0.38\\
    &&&&&  1.00 &  -0.97\\
    &&&&&&  1.00\end{bmatrix*}$ } \\ 
    $\gamma_{\SLJ{3}{P}{2}}^{(0)} =$ & $(61.98 \pm 7.54 ) \cdot a_t^2$ & \\
    $\gamma_{\SLJ{3}{D}{2}}^{(0)} =$ & $(-1785.4 \pm 726.4 ) \cdot a_t^4$ & \\
    $\gamma_{\SLJ{3}{P}{0}}^{(0)} =$ & $(153.2 \pm 12.6 ) \cdot a_t^2$ & \\
    $\gamma_{\SLJ{3}{D}{1}}^{(0)} =$ & $(1057 \pm 438 ) \cdot a_t^4$ & \\
    $\gamma_{\SLJ{3}{S}{1}}^{(0)} =$ & $(7.91 \pm 0.65 )$ & \\
    $\gamma_{\SLJ{3}{S}{1}}^{(1)} =$ & $(-214.04 \pm 28.91 ) \cdot a_t^2$ & \\[1.3ex]
    &\multicolumn{2}{l}{ $\chi^2/ N_\mathrm{dof} = \frac{48.63}{36-7} = 1.68$\,,}
    \end{tabular}
\label{eq:refluscher}
\end{equation}
where the quoted error is purely statistical. 
In this case, where $g=0$, the $\gamma$ parameters of the physical K matrix as defined in Ref.~\cite{Whyte:2024ihh} coincide with $\lhcgamma$ from the definition of $\Kpar_0$ in Eq.~\eqref{eq:k_poly}.
When the initial and final state labels coincide, we will make use of the simplified notation $\gamma_{ii}^{(n)} \to \gamma_{i}^{(n)}$.
The matrix in square brackets shows the statistical correlations between the resulting parameters.

We note that all the allowed partial waves up to $\ell=2$ in Table~\ref{tab:subduction} are necessary to minimally describe the spectrum considered, resulting in $\chi^2/ N_\mathrm{dof} = 1.68$.
In this case, some of the parameters have considerable uncertainties, namely all the $D$-wave parameters have roughly $50\%$ relative statistical errors.

By setting $g=0$, this parametrization corresponds to the physical K-matrix, and thus one can directly consider the analytical continuation of the corresponding scattering amplitude into the second sheet.
There, one finds a virtual bound state at $\sqrt{s} \approx 0.683 \, a_t^{-1} \approx 3870~{\rm MeV}$ in the $J^P=1^+$, $S$-wave amplitude, consistent with the $T_{cc}^+$ pole found in Ref.~\cite{Whyte:2024ihh}.
It is also good to note that a resonance pole is found at $\sqrt{s} \approx ( 0.697 + \frac i2 \, 0.034 ) \, a_t^{-1} \approx (3950 + \frac i2\, 190)~{\rm MeV}$ in the $0^-$ $P$-wave amplitude, which was similarly found in Ref.~\cite{Whyte:2024ihh} even though not consistently across the K-matrix parametrization variations.

\begin{figure}[htbp]
    \centering
    \includegraphics[width=0.9\textwidth]{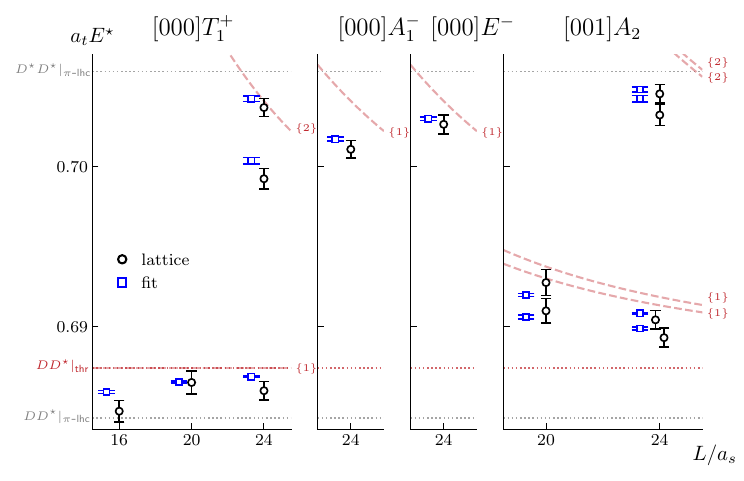}
    \includegraphics[width=0.9\textwidth]{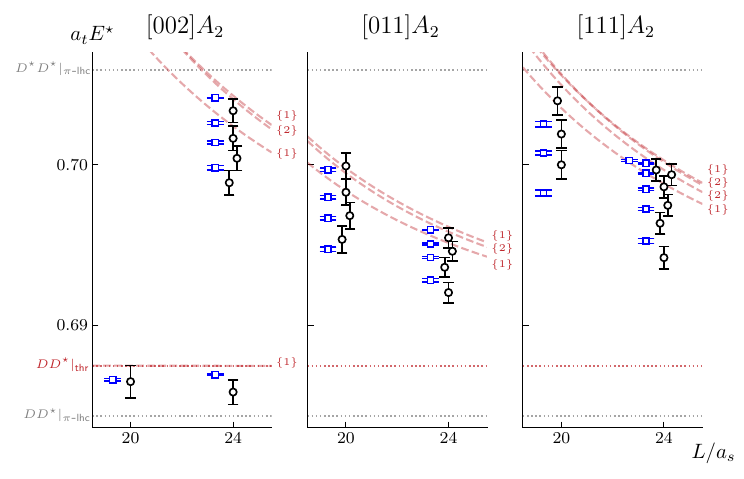}
    \caption{
    Lattice energies computed in Ref.~\cite{Whyte:2024ihh} (black circles) and the spectrum derived from the parametrization in Eq.~\eqref{eq:reflhc} as a function of $L/a_s$ (blue squares) on each of the irreps considered.
    The errorbars represent only statistical uncertainties derived from the quoted $\Kpar_0$ parameters and $g$ coupling.
    The data points were horizontally displaced for clarity.
    Noninteracting energies are represented by dashed curves (red), while the $DD^\star$ kinematic threshold (red) and left-hand branch points (gray) are shown as horizontal dashed dotted lines, respectively.
    The degeneracies of the noninteracting levels are indicated in curly brackets.
    }
    \label{fig:specvsL_all}
\end{figure}

\subsection{Reference parametrization for $g$ as a fit parameter}
\label{sec:reflhc}

Here we use the quantization condition in Eq.~\eqref{eq:QC_cc_spin} to simultaneously constrain the $D^\star D\pi$ coupling $g$ and the parameters of $\Kpar_{0}$ from finite-volume energies, allowing both to vary during the minimization of the spectrum $\chi^2$ defined in Ref.~\cite{Wilson:2014cna}.
We fix the mass of the exchange particle to that of the pion mass obtained from the lattices used to extract the finite-volume energies, i.e. $m_\pi =0.06906 \, a_t^{-1} \approx 391$~MeV.

We choose the reference parametrization to be a coupling $g$ and a $\Kpar_0$-matrix polynomial given by
\begin{equation}
\renewcommand{\arraystretch}{1.5}
    \begin{tabular}{rll}
    $\lhcgamma_{\SLJ{3}{P}{2}}^{(0)} =$ & $(37.2 \pm 5.0) \cdot a_t^2$ & \multirow{5}{*}{ $\begin{bmatrix*}[r]   1.00 &   0.57 &   0.33 &  -0.27 &   0.58\\
    &  1.00 &   0.18 &  -0.24 &   0.85\\
    &&  1.00 &  -0.88 &   0.14\\
    &&&  1.00 &  -0.19\\
    &&&&  1.00\end{bmatrix*}$ } \\ 
    $\lhcgamma_{\SLJ{3}{D}{1}}^{(0)} =$ & $(927 \pm 507) \cdot a_t^4$ & \\
    $\lhcgamma_{\SLJ{3}{S}{1}}^{(0)} =$ & $(3.40 \pm 0.41)$ & \\
    $\lhcgamma_{\SLJ{3}{S}{1}}^{(1)} =$ & $(-41 \pm 25) \cdot a_t^2$ & \\
    $ g = $ & $(12.57 \pm 0.33)$ & \\[1.3ex]
    &\multicolumn{2}{l}{ $\chi^2/ N_\mathrm{dof} = \frac{27.88}{36-5} = 0.90$\,,}
    \end{tabular}
    \label{eq:reflhc}
\end{equation}
where $g$ was statistically treated in the same way as the $\lhcgamma$ parameters.
Here, the value of the cutoff-function parameter entering Eq.~\eqref{eq:Hfunc} is $\hat\alpha 170$ in lattice units, based on a goodness-of-fit criterion as explained in Secs.~\ref{sec:alphavariation} and ~\ref{sec:par_vars}.
In Fig.~\eqref{fig:specvsL_all}, the finite-volume spectrum derived from this parametrization is shown together with the lattice data.

It is noticeable that the $\chi^2/ N_\mathrm{dof} = 0.90$ is considerably lower than the result for fixed $g = 0$ in Eq.~\eqref{eq:refluscher}.
This is the case even though only four $\Kpar_0$ parameters are needed, suggesting that the exchange coupling $g$ effectively absorbs the contribution from the additional parameters in Eq.~\eqref{eq:refluscher}, namely from $\SLJ{3}{P}{0}, \SLJ{3}{D}{2}$ and $\SLJ{3}{D}{3}$ partial waves.
This points to a more effective description of the data when accounting for left-hand cut effects, even if the lowest finite-volume energy is just above the left-hand branch point (see Fig.~\ref{fig:specvsL_all}).
We leave further comments to Sec.~\ref{sec:par_vars}, where we show all parametrization forms considered.

\subsection{Choice of the Parameter $\alpha$}
\label{sec:alphavariation}

We study the sensitivity of the result above to the parameter $\alpha$ appearing in the cutoff function defined in \eqref{eq:Hfunc}.
Given the formalism presented in Ref.~\cite{Raposo:2025dkb}, this parameter must be nonzero.
This is because all momentum integrals and sums appearing in the formalism are divergent in the $\alpha\to 0$ limit.
On the other hand, if one fixes $\alpha$ to a large value, one runs the risk of introducing enhanced finite-volume artifacts.
As a result, it is expected that $\alpha$ needs to be moderately small but nonzero. 

It is important to emphasize that by fixing $\alpha$ to a specific value, we have chosen a scheme within which to define $\Kc_0$.
As long as the same scheme is used for both the finite-volume quantization condition and the infinite-volume integral equations, one is assured to have amplitudes that are independent of this scheme.

The parameter $\alpha$ is dimensionful with units of $\rm [MeV]^{-2}$.
Iit is useful to introduce a corresponding dimensionful variable that can be interpreted as a cutoff scale in units of $\rm [MeV]$
\begin{equation}
    \Lambda_\alpha \equiv 1/\sqrt{\alpha} \,.
\end{equation}
As a consequence, the contribution of momenta for which $k^\star \gg \Lambda_\alpha$ will be exponentially suppressed.
One might be tempted to make $\Lambda_\alpha$ proportional to $2\pi/L$, since it is a momentum scale.
This would then make the cutoff scale volume-dependent and also make the matching with the infinite-volume formalism impractical.
Instead, we define $\alpha$ in lattice units, 
\begin{equation}
    \hat \alpha \equiv \alpha a_t^{-2} \,,
\end{equation}
and it is $\hat \alpha$ that we keep fixed to the values given below. 

In this calculation, we are interested in the momentum range between $| q^\star | \lesssim 450{\rm MeV}$ corresponding to the region between just below the $D D^\star$ single-pion left-hand cut and the $D^\star D^\star$ threshold. 
Requiring that $| q^\star | \lesssim \Lambda_\alpha$, we pick the three values of $\hat\alpha$ listed in Table~\ref{tab:alpha_scan}.

\begin{table}[htbp]
    \centering
    \renewcommand{\arraystretch}{1.5}
    \setlength{\tabcolsep}{12pt}
    \begin{tabular}{|c||ccc|}
    \hline
    $\hat\alpha = \alpha a_t^{-2}$
    & 170 & 110 & 50 \\
    \hline
    $\Lambda_\alpha\,[\mathrm{GeV}]$
    & 0.435 & 0.540 & 0.801 \\
    \hline
    \end{tabular}
    \caption{
    Choices of cutoff scales $\Lambda_\alpha = 1/\sqrt{\alpha}$ explored in this work, defined by the cutoff-function parameter $\alpha$.
    }
    \label{tab:alpha_scan}
\end{table}

In this analysis, we explore the range of values of $\hat\alpha$ given above, and then use weights based on the Akaike information criterion (AIC) as a guide as to which values are likely to be statistically optimal.
We employ weights proportional to $\exp(- \frac12 {\rm AIC})$, where for a given fit, ${\rm AIC} = \chi^2 + 2 N_{\rm pars}$ and $N_{\rm pars}$ is the number of parameters~\cite{Jay:2020jkz}.
When using the opacity to represent the AIC-preferred variations in plots, we normalize each weight by the largest value, namely the weight of the reference fit.
In summary, we find empirical evidence of the values of $\hat\alpha$ that are statistically preferred.

\subsection{Sensitivity to the Momentum Cutoff $\pcut$}
\label{sec:pcutvariation}

In practice, it is also necessary to introduce a hard momentum cutoff, $\pcut$, for each given value of $\hat{\alpha}$.
This is to render the sums defining the finite-volume function $\Cc_L$, defined in \eqref{eq:CL_spin}, finite.
This will, in turn, also truncate the momenta appearing in the integrals defining $\Mc_\Ec$, Eq.~\eqref{eq:Mt_int_spin}, and other subsequent infinite-volume functions, finite.
For the system considered here, the main computational bottleneck is the evaluation of $\Cc_L$, since this is being determined ``\emph{on the fly}'' while the OPE coupling $g$ is being constrained by the lattice spectrum. 

In practice, we pick a range of reasonable values of $\pcut$, test the convergence with this variable, and report the results for the largest value of $\pcut$.
Here, we report some guidance we used to choose reasonable minimal values of this parameter that may help the reader. 

Having chosen a reasonable value of $\hat{\alpha}$, which, as we explained above, should be constrained by the physical range of $q^\star$, one can then fix $\pcut$ to introduce errors that scale at worst as the already ignored exponentially suppressed finite volume errors, namely $e^{-m_\pi L}\lesssim e^{-4}\sim 0.02 $. 
This gives the simple criterion, 
\begin{equation}
    \exp ( -\alpha \pcut^2 ) \leq 0.02 \,.
    \label{eq:pcut}
\end{equation}
We can further refine this argument and thereby have a better intuition of the parametric behavior of this residual error.
If we assume $\pcut^2 \gg q^{\star 2}$, we can estimate the error of truncating a generic sum for any angular momenta.
If we look at the tail of momenta that have been ignored in the sum, one can safely replace the residual sum with an integral, up to negligible exponentially suppressed errors.
Doing this for $S$ and $D$ partial waves, one gets an error that scales as
\begin{align}
    I_{SS}(\pcut,\alpha)
    &\equiv
    \int_{\pcut}^{\infty} dp\, e^{-\alpha p^2}
    =
    \frac{\sqrt{\pi}}{2\sqrt{\alpha}}\,
    \operatorname{erfc}\!\left(\pcut\sqrt{\alpha}\right) \,,
    \label{eq:pcut_tail_s}\\
        I_{SD}(\pcut,\alpha;q)
    &\equiv
    \int_{\pcut}^{\infty} dp\,
    \left(\frac{p}{q}\right)^2 e^{-\alpha p^2}
    =
    \frac{1}{q^2}
    \left[
    \frac{e^{-\alpha \pcut^2}\pcut}{2\alpha}
    +
    \frac{\sqrt{\pi}}{4\alpha^{3/2}}\,
    \operatorname{erfc}\!\left(\pcut\sqrt{\alpha}\right)
    \right],
    \label{eq:pcut_tail_sd}
    \\
    I_{DD}(\pcut,\alpha;q)
    &\equiv
    \int_{\pcut}^{\infty} dp\,
    \left(\frac{p}{q}\right)^4 e^{-\alpha p^2}
    =
    \frac{1}{q^4}
    \left[
    \frac{e^{-\alpha \pcut^2}\pcut\left(3+2\alpha \pcut^2\right)}{4\alpha^2}
    +
    \frac{3\sqrt{\pi}}{8\alpha^{5/2}}\,
    \operatorname{erfc}\!\left(\pcut\sqrt{\alpha}\right)
    \right].
    \label{eq:pcut_tail_dd}
\end{align}
Although this is not the exact form of the error on $\Cc$ and $\Cc_L$, it does give us some insight into its expected asymptotic behavior.
For starters, we see that if we fix $\alpha \pcut^2$, the error will grow inversely with $\alpha$ to some power.
This is consistent with what we already observed, namely that the limit $\alpha \to 0$ is not well defined.
Here, this manifests itself in the fact that the smaller $\alpha$ is, the larger the tail one will need to integrate.
Also, as one would expect, the barrier factors of the higher partial waves increase this effect. 

Using a larger $\alpha$ might lead one to assume that one can use smaller values of $\pcut$.
That is indeed the case, but one must remember that this is still required to satisfy $\pcut \gtrsim |q^{\star }| $.
In lattice units, the kinematics considered here are fixed by $a_t^2q^{\star 2} \lesssim 0.006$, and thus we restrict our attention to $a_t^2 \pcut^2 \gtrsim 0.015$.
As previously stated, in practice, we simply vary $\pcut$ in our analysis and verify the results are not sensitive to the choice of $\pcut$.

As a last remark, note that in a finite volume with a spatial extent $L$, the momentum cutoff $p_{{\rm cut}, L}$ can only assume discrete values corresponding to the available finite-volume momentum shells.
Across the three volumes $L/a_s \in \{16,20,24\}$ used to constrain $\Kpar_0$ and the coupling $g$ for a given $\alpha$, we round up $p_{{\rm cut}, L}$ to the next available value in the volume.
This ensures that $\Kpar_0$ and $g$ are volume-independent up to a target residual error.
Having explored a range of $p_{{\rm cut}}$ and found no dependence of our result on this variable, we report the results obtained using the largest value of this variable.

\begin{table*}[!ht]
    \centering
    \setlength{\tabcolsep}{0.75em}{\renewcommand{\arraystretch}{1.85}{%
    \small
    \begin{tabular}{|c|c|c|c|c||cc|cc|cc|c}
        \hline
        \multirow{2}{*}{\#} &
        \multicolumn{2}{c|}{$P=+$} &
        \multicolumn{2}{c||}{$P=-$} &
        \multicolumn{2}{c|}{$\hat\alpha=170$} &
        \multicolumn{2}{c|}{$\hat\alpha=110$} &
        \multicolumn{2}{c|}{$\hat\alpha=50$} \\
        \cline{2-11}
        &
        $\SLJ{3}{S}{1}$ &
        $\SLJ{3}{D}{1}$ &
        $^3P_2$ &
        $^3P_0$ &
        $g$ &
        $\chi^2/N_\mathrm{dof}$ &
        $g$ &
        $\chi^2/N_\mathrm{dof}$ &
        $g$ &
        $\chi^2/N_\mathrm{dof}$ \\
        \hline
    
        \multicolumn{11}{|c|}{$\Kpar_0$ polynomial} \\
        \hline\hline
    
        $1$ &
        $0,1$ & -- & $0$ & -- &
        $12.01(18)$ &
        $\frac{33.93}{36-4}=1.06$ &
        $11.53(16)$ &
        $\frac{32.34}{36-4}=1.01$ &
        $10.76(13)$ &
        $\frac{31.19}{36-4}=0.97$ \\
    
        $2$ &
        $0$ & $0$ & $0$ & -- &
        $12.44(31)$ &
        $\frac{30.54}{36-4}=0.95$ &
        $11.82(27)$ &
        $\frac{31.74}{36-4}=0.99$ &
        $10.92(19)$ &
        $\frac{33.37}{36-4}=1.04$ \\
    
        $3$ &
        $0,1$ & $0$ & $0$ & -- &
        $\mathbf{12.57(33)}$ &
        $\mathbf{\frac{27.88}{36-5}=0.90}$ &
        $11.91(28)$ &
        $\frac{27.98}{36-5}=0.90$ &
        $11.13(13)$ &
        $\frac{30.36}{36-5}=0.98$ \\
    
        $4$ &
        $0$ & -- & $0$ & -- &
        $12.01(18)$ &
        $\frac{35.30}{36-3}=1.07$ &
        $11.52(16)$ &
        $\frac{35.02}{36-3}=1.06$ &
        $10.74(13)$ &
        $\frac{36.20}{36-3}=1.10$ \\
        \hline
    
        \multicolumn{11}{|c|}{$\Kpar_0^{-1}$ polynomial} \\
        \hline\hline
    
        $5$ &
        $0,1$ & -- & $0$ & $0$ &
        $11.11(70)$ &
        $\frac{29.26}{36-5}=0.94$ &
        $11.22(60)$ &
        $\frac{28.90}{36-5}=0.93$ & --
        & -- \\
    
        $6$ &
        $0$ & $0$ & $0$ & $0$ &
        $12.44(31)$ &
        $\frac{30.54}{36-5}=0.99$ &
        $12.0(10)$ &
        $\frac{31.68}{36-5}=1.02$ & --
        & -- \\
        \hline
    \end{tabular}
    \caption{\label{tab:parvar}
    Variations of the $\Kpar_0$ parametrization from fits to finite-volume energies below the $D^\star D^\star$ left-hand cut, with $g$ treated as a free parameter.
    The entries of the table display the values of $n$ for which $\lhcgamma_{ij}^{(n)}$ or $\lhcinvkc_{ij}^{(n)}$ from Eqs.~\eqref{eq:k_poly},~\eqref{eq:kinv_poly} are nonzero.
    The amplitudes are grouped by parity $P$, and the ones labeled by a single partial wave correspond to diagonal amplitudes.
    The ``--'' or the absence of a partial wave means that the corresponding parameters are fixed to zero.
    The $\hat\alpha$ columns denote the variations of each $\Kpar_0$ form for $\hat\alpha \in \{50,110,170\}$, with the corresponding $DD^\star\pi$ coupling $g$ and statistical uncertainties, and $\chi^2/ N_\mathrm{dof}$ given in nested columns.
    The variation denoted in bold is the reference parametrization, quoted in Eq.~\eqref{eq:reflhc} and represented in Fig.~\ref{fig:Tcc_rhoM_abs2_var} together with all variations.
    The ``--'' under $\hat\alpha$ columns correspond to parametrizations that did not result in a reasonable fit.}
    }}%
\end{table*}

\subsection{Parametrization Variations}
\label{sec:par_vars}

We assess the dependence of the results on the form of the $\Kpar_0$ parametrization by employing variations based on Eqs.~\eqref{eq:kinv_poly} and \eqref{eq:k_poly} and the possible partial waves listed in Table~\ref{tab:subduction}.
We also assess the dependence on the dimensionless cutoff parameter $\hat\alpha$, which in the current analysis we keep fixed to the values $\hat\alpha \in \{50,110,170\}$.
We show in Table~\ref{tab:parvar} the subset of parametrizations that reasonably describe the spectrum.

We observe an overall deterioration in the goodness of fit for $\hat\alpha < 110$, which corresponds to the particularly faint AIC-weighted results in Fig.~\ref{fig:pole}.
Nevertheless, we observe that the exchange coupling $g$ remains consistent within uncertainties across the explored values of $\hat\alpha$.
For the reference parametrization, we also check that the results for $\hat \alpha > 170$ are less favored by the AIC weighting, as expected.

Beyond the variations in Table~\ref{tab:parvar}, we have also examined the inclusion of other $\Kpar_0$ parameters present in the Lüscher-only result Eq.~\eqref{eq:refluscher}.
In particular, the constant parameters in $J^P = 0^-, 2^\pm, 3^+$ partial waves are largely consistent with zero and their inclusion does not improve the $\chi^2 / N_\mathrm{dof}$.
Instead, their introduction generates high statistical correlations to other $\Kpar_0$ parameters and $g$, which prevents a well-behaved minimization of the $\chi^2$ function.

We report the results from $\Kpar_0^{-1}$ parametrizations to provide a nontrivial crosscheck to the results from $\Kpar_0$ polynomials.
The $\Kpar_0^{-1}$ polynomial parametrizations need a nonzero $\SLJ{3}{P}{0}$ constant term in order to minimally function because of their particular form (see Eq.~\ref{eq:kinv_poly}), although this term appears consistent with zero and very correlated to other parameters.
The value of the exchange coupling $g$ is consistent across all variations.

\begin{figure}[htbp]
    \centering
    \includegraphics[width=.80\textwidth]{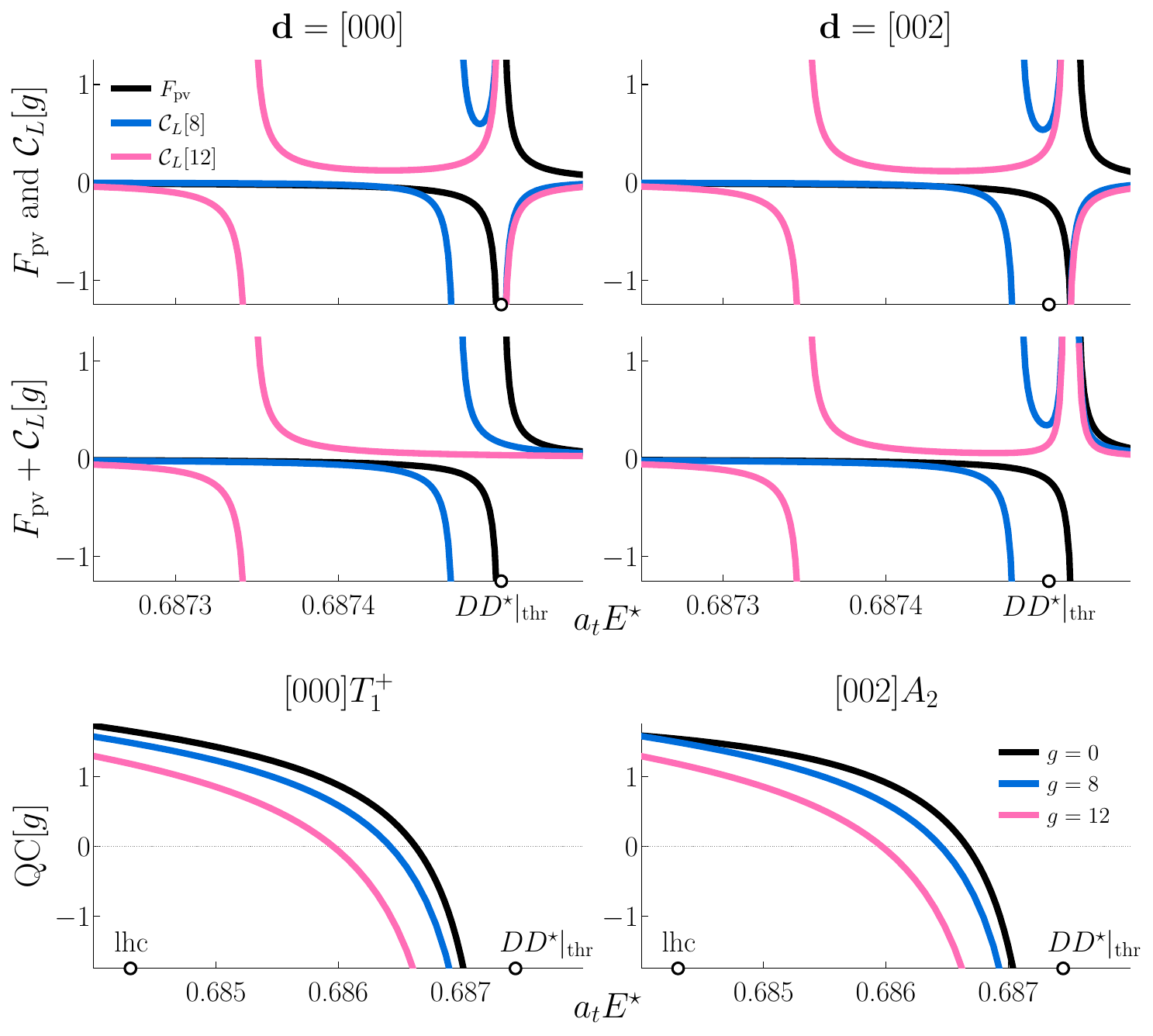}
     \caption{
     In the top panel, we show the values of $(\Fpv )_{\ell'm' ; \ell m}$ and $(\Cc_{L}[g])_{\ell'm'm_S' ; \ell mm_S}$ for $\ell'm' ; \ell m = 10;10$ and $m_S'=m_S=1$, where we vary $g \in \{8,12\}$ and fix the spatial extension to be $L/a_s = 24$.
     The left and right columns correspond to $\svec{d} = [000]$ and $[002]$, respectively.
     In the middle panel, we show the sum of $\Fpv$ and $\Cc_{L}[g]$ at the same values of $g$.
     In the bottom panel, we show a plot of the quantization condition in the real-valued form $\mathrm{QC}[g] \equiv 1 + \Kc_0 (\Fpv + \Cc_L)$ as a function of energy for $g\in\{0,8,12\}$, where the $\Kc_0$ matrix is chosen as the K-matrix with fixed $g=0$ from Eq.~\eqref{eq:refluscher}.
     The values of $\alpha$ and $\pcut$ are held fixed to representative values across all the curves.
     }
    \label{fig:1}
\end{figure}

\subsection{OPE-dependence of the finite-volume spectrum}

Given that this is the first implementation of this formalism in the analysis of the spectrum, we provide some qualitative examples on how the quantization condition depends on the value of the exchange coupling $g$ entering the OPE.

First, let us consider the building blocks that appear in the quantization condition, Eq.~\eqref{eq:QC_cc_spin}, namely $\Fpv$ and $\Cc_L$.
For simplicity, we only show the quantization condition components with dominant contribution to the $S$-wave of $\Kc_0^{1^+}$, and focus on the region between the left-hand cut and the $DD^\star$ threshold.
We will consider the effect of this for the $T_1^+$ and the $[002]A_2$ irreps at first, where finite-volume energy levels are expected to be present just below threshold.

We compare $\Cc_L$ and $\Fpv+\Cc_L$ at different values of the exchange coupling $g$ within a moderate range.
In the top panel of Fig.~\ref{fig:1}, we observe a characteristic ``double-pole'' structure in $\Cc_L[g]$ for $g>0$; the first one near the $DD^\star$ threshold and a second one appearing at lower energies as $g$ is increased.
The $\Cc_L$ pole near threshold has the opposite sign to that of $\Fpv$ at the same position, the latter corresponding to a noninteracting level.
Those poles thus cancel each other in $\Fpv + \Cc_L$, as shown in the middle panel of Fig.~\ref{fig:1}.
This cancellation occurs exactly on top of the threshold when the system is at rest, and slightly above it for an unequal-mass system with overall momentum $\mathbf d = [002]$.
This leads to a function where the second pole appears increasingly displaced toward lower energies when compared to the pole from $\Fpv$.

Having studied the building blocks, we can now proceed to look at the quantization condition.
For this, we will need a parametrization of $\Kpar_0$, for which we use the $g=0$ result from Eq.~\eqref{eq:refluscher}.

In the lowest panel of Fig.~\ref{fig:1}, we plot the quantization condition in a region where contributions from the $S$-wave amplitude in Eq.~\eqref{eq:refluscher}  are expected to dominate the energy shifts for $g=0$.
The colorful lines correspond to the quantization condition for different fixed values of $g\in\{8,12\}$ while keeping $\Kpar_0$ fixed to that of Eq.~\eqref{eq:refluscher}.
The increasing values of $g$ change the position of the quantization condition zeroes, which in this specific case imply energy levels predicted further away from threshold towards the left-hand cut.
We note that the overall number of zero crossings stays the same as $g$ is increased, and thus the level counting is conserved.
\\

\section{Scattering amplitudes and poles}
\label{sec:scatamps_poles}

Before discussing the analytic continuation of the amplitudes into the second Riemann sheet, we briefly summarize the key ingredients needed to obtain the amplitude on the physical Riemann sheet.
First, we evaluate the partial-wave-projected OPE kernels described in Secs.~\ref{sec:OPE_1plus} and \ref{sec:OPE_pwaves}; an on-shell illustration for all partial waves considered here is shown in Fig.~\ref{fig:OPE}.
We then solve the integral equation for $\Mc_\Ec$, Eq.~\eqref{eq:Mt_int_spin}.
For $J^P=1^+$ ,this is a coupled $\SLJ{3}{S}{1}$-$\SLJ{3}{D}{1}$ equation, while the $P$ waves considered here are single-channel equations.
In practice, the momentum integrals are discretized with Gauss-Legendre quadrature, and the equations are solved as finite matrix equations in the combined momentum and channel space, following the methods of Refs.~\cite{Dawid:2023jrj,Briceno:2024ehy, Briceno:2025yuq} for three-body systems. 

Once $\Mc_\Ec$ is known, the remaining OPE-induced quantities $\Rc$, $\Lc$, and $\Cc$ are obtained from Eqs.~\eqref{eq:iR}, \eqref{eq:iL}, and \eqref{eq:C}.
Combining these with $\Kc_0$ gives the full physical-sheet amplitude through Eq.~\eqref{eq:main_amp}.
Above threshold, we use the standard pole-deformed physical-sheet contour to avoid the pole in $\Delta_{2,\infty}$ at the on-shell momentum, thereby improving the convergence of the solutions with respect to the number of points used in the contour.
This gives partly on-shell amplitudes $\Mc_\Ec(p',q^\star)$ and, by one final use of the integral equation, the fully on-shell quantity $\Mc_\Ec(q^\star,q^\star)$.

\begin{figure}[htbp]
    \centering
    \includegraphics[width=0.96\textwidth]{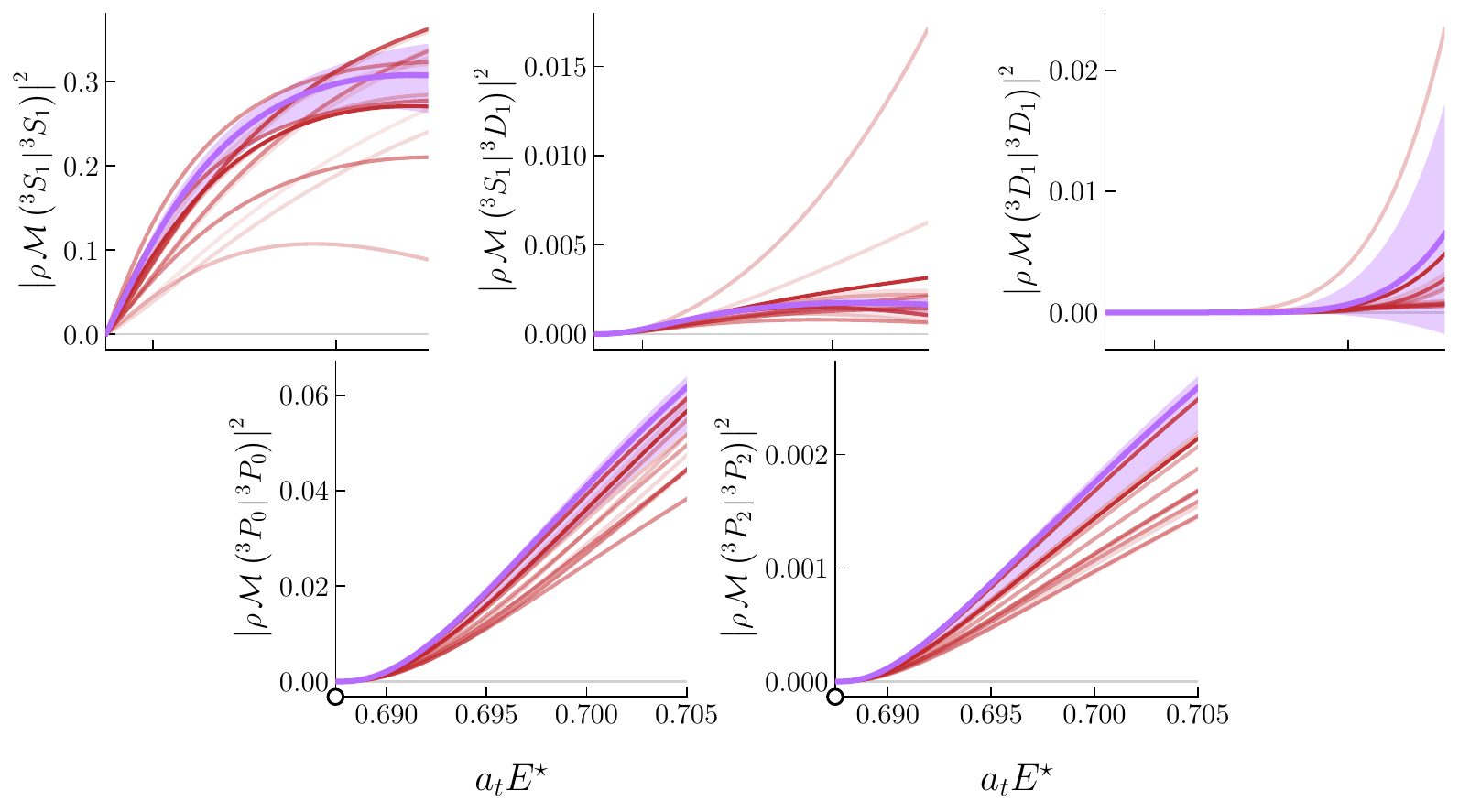}
    \caption{
    Shown are the values of $|\rho\,\Mc|^2$ above threshold for all amplitudes determined.
    The reference fit is shown with statistical uncertainty bands.
    The additional curves show central values from the parametrization variations in Table~\ref{tab:parvar}, with opacity set by the relative AIC weight in the same convention as Fig.~\ref{fig:pole}.
    The left column shows the coupled $J^P=1^+$ components, while the right column shows the $P$-wave amplitudes. 
    }
    \label{fig:Tcc_rhoM_abs2_var}
\end{figure}
\begin{figure}[htbp]
    \centering
    \includegraphics[width=.7\textwidth]{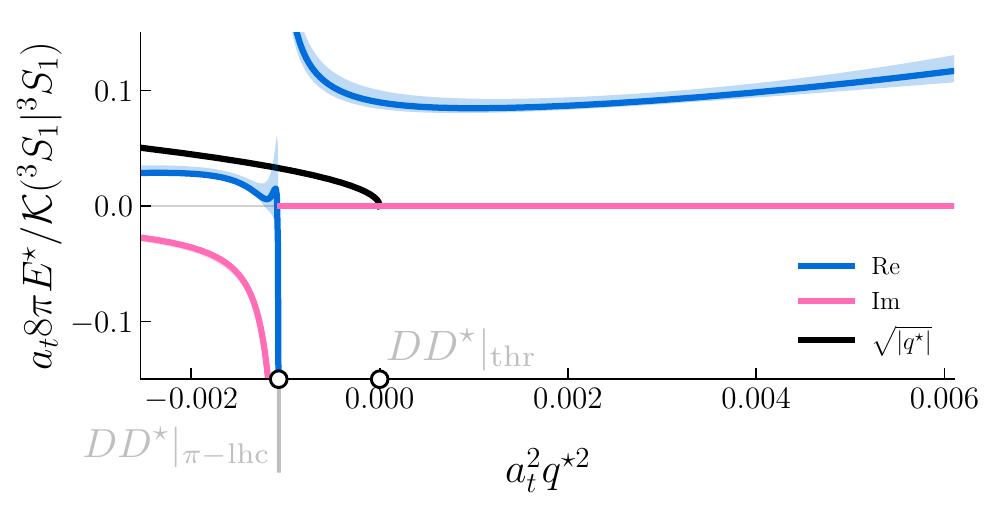}
     \caption{
     Shown is  $a_t 8\pi E^\star / \Kc(\SLJ{3}{S}{1}|\SLJ{3}{S}{1})$ as a function of $a_t^2 q^{\star 2} $, where $\Kc$ has been defined via $\Kc^{-1}=\Mc^{-1}+i\rho$.
     This quantity reduces to $a_t q^\star\cot\delta$ in the absence of $\SLJ{3}{D}{1}$ mixing and left-hand-cut effects.   
     }
    \label{fig:Tcc_pcot}
\end{figure}

\clearpage 
Figure~\ref{fig:Tcc_rhoM_abs2_var} shows the above-threshold quantity $|\rho\,\Mc|^2$ for all the parametrization variations.
This is a particularly useful quantity to consider, given that it is bounded by $1$ for all components above threshold. 
The bands show the statistical uncertainty of the reference parametrization, while the additional curves show the central values obtained from all the alternative $\Kpar_0$ and $\Kpar_0^{-1}$ polynomials listed in Table~\ref{tab:parvar}.
The opacity of each curve reflects its likelihood given by its corresponding relative AIC weight. 

In Fig.~\ref{fig:Tcc_pcot}, we show $a_t 8\pi E^\star / \Kc(\SLJ{3}{S}{1}|\SLJ{3}{S}{1})$, where we have defined $\Kc^{-1} = \Mc^{-1}+i\rho$.
This definition does not require $\Kc$ to be real along the left-hand cut.
In the limit of no mixing with the $\SLJ{3}{D}{1}$ channel, this would be equal to the usual $(a_t q^\star) \cot\delta$.
We see that below threshold this does not cross $\sqrt{|q^\star|}$, as it would be required for a virtual bound state.
This is due to the presence of the left-hand cut.

\begin{figure*}[b]
    \centering
    \includegraphics[width=.75\textwidth]{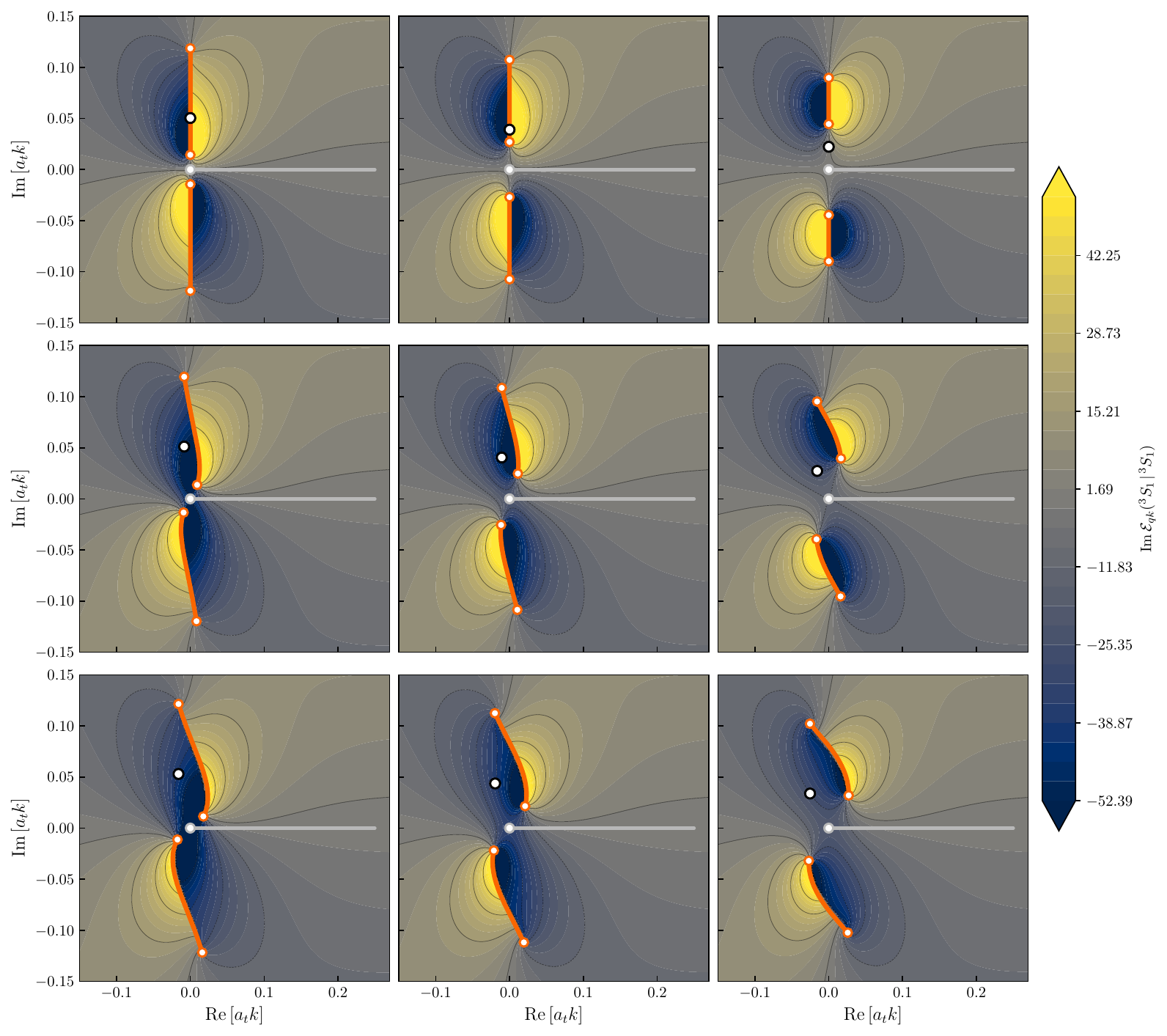}
    \caption{
    Imaginary part of a partly on-shell OPE, $\Ec(q^\star, k)$, kernel in the complex integration-momentum plane for the reference $\Kc_0$ and $g$.
    The panels scan the same complex energies used in the analytic continuation checks.
    The gray line denotes the integration contour, the open black circles denote the on-shell momentum, and the orange lines denote the OPE cuts associated with $\zeta(k,q^\star)=x$ for $x\in[-1,1]$.
    The displayed energies are chosen below threshold and probe the region relevant to the analytic continuation.
    }
    \label{fig:ope_qk_reference}
\end{figure*}

\subsection{Analytic continuation and checks}
\label{sec:analytic_continuation_checks}

The pole determinations quoted in the main text require evaluating the same integral equations at complex energies.
Since the OPE kernel has logarithmic branch points, this continuation is meaningful only if the external on-shell point and the momentum-integration contour remain in a common analytic domain and can be connected without crossing the OPE cuts.
This criterion follows the analysis of Refs.~\cite{Brayshaw:1968yia, Dawid:2023jrj,Dawid:2023kxu}.
Here we summarize the checks performed for the kinematics relevant to the $T_{cc}^+$ pole.

The basic momentum-space structure of the partly on-shell OPE, $\Ec(q^\star, k)$, which plays a critical role in the integral equation given by Eq.~\eqref{eq:Mt_int_spin}, is shown in Fig.~\ref{fig:ope_qk_reference}.
For real energies below threshold, the on-shell momentum lies on the imaginary axis, while the integration contour can remain on the real axis.
The orange curves show the OPE branch cuts, which are obtained by solving
\begin{equation}
    \zeta(k,q^\star)=x,\qquad x\in[-1,1],
\end{equation}
with $\zeta$ defined in Eq.~\eqref{eq:zeta}.

\begin{figure*}[b]
    \centering
    \includegraphics[width=.75\textwidth]{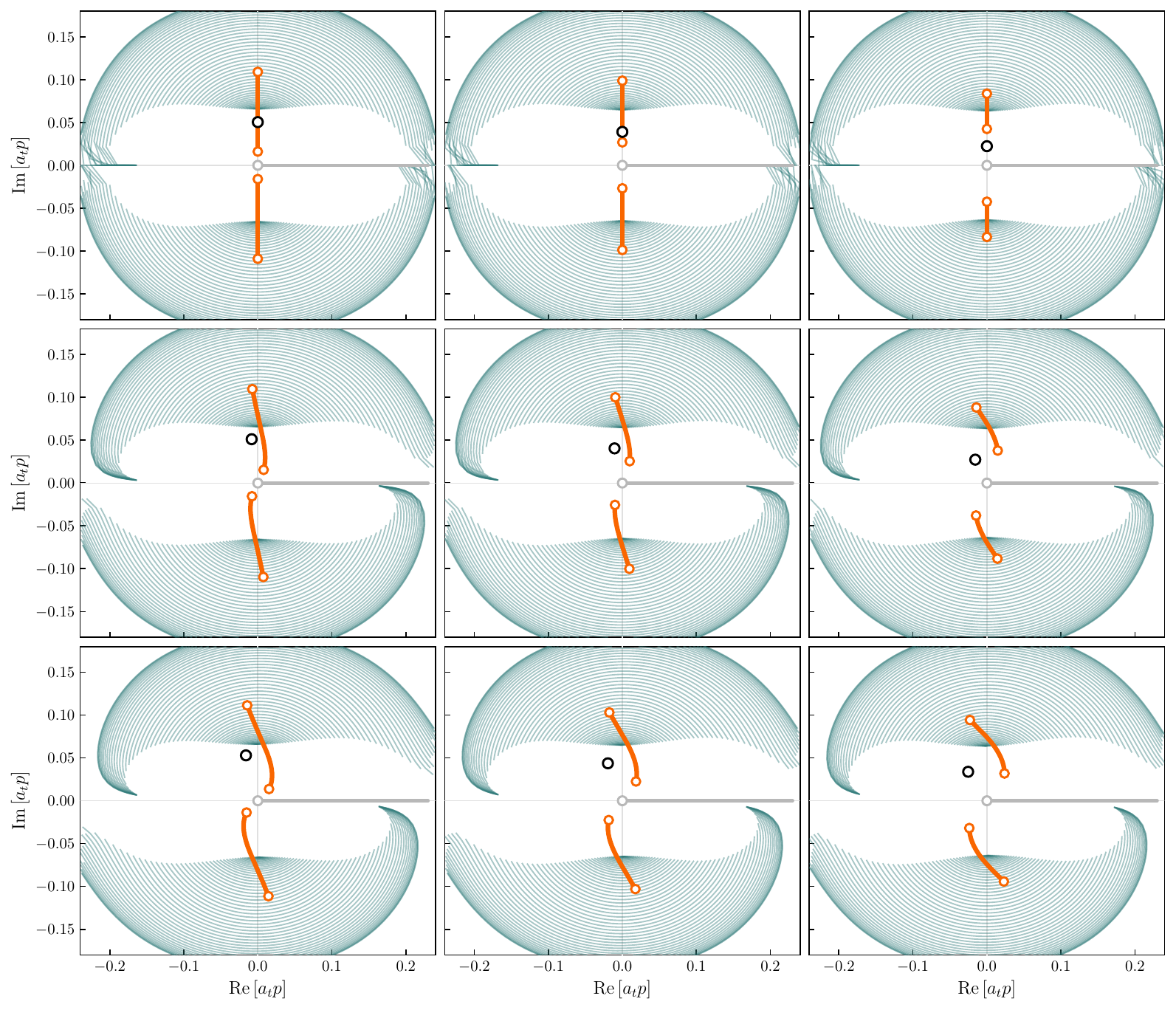}
    \caption{
    Domain-of-non-analyticity check for the OPE kernel in the
    complex momentum plane, following the criteria of
    Ref.~\cite{Dawid:2023jrj}.  The blue curves are obtained by varying the
    spectator momentum up to the cutoff and the angular variable over its
    physical interval.  The numerical integration contour and on-shell
    point remains outside the obstructing region for the energies relevant
    to the pole.
    }
    \label{fig:ope_nonanalyticity_domain}
\end{figure*}

\begin{figure}[b]
    \centering
    \includegraphics[width=.6\textwidth]{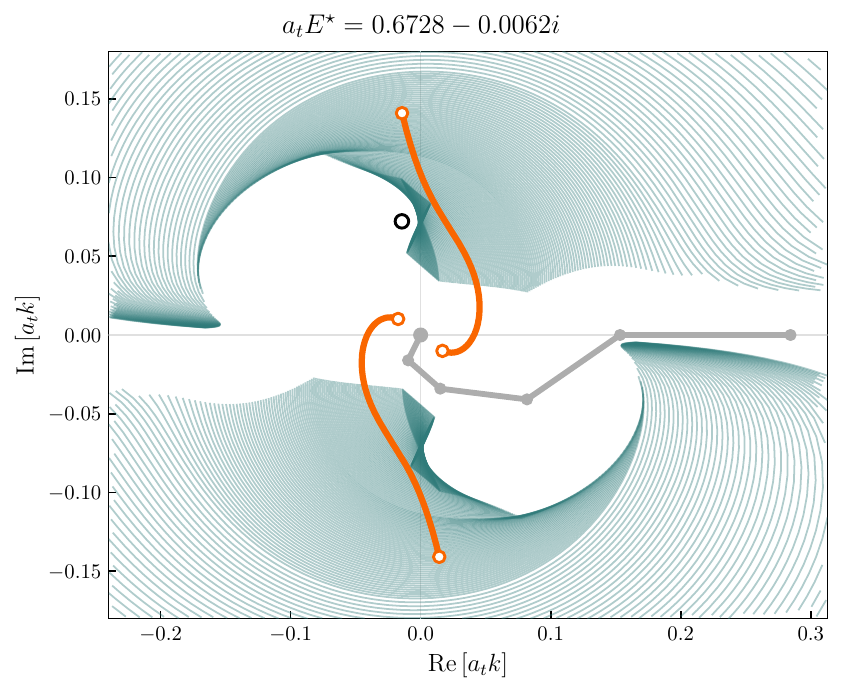}
    \caption{Same as Fig.~\ref{fig:ope_nonanalyticity_domain}, but for the deformed contour used for the $\Kc_0$ parametrizations with $\hat\alpha=50$.}
    \label{fig:contour_v2}
\end{figure}

To make this more explicit, let us consider two possible singularity structures that can appear when we make the symmetrization appearing in Eq.~\eqref{eq:Ec_spin} explicit, in the scenario where the rightmost particle is on-shell.
This can be done most conveniently by introducing two expressions for the Mandelstam variable, $u$, that agree in the fully on-shell limit,
\begin{align}
u_1(k,q^\star,x;s)
&=
\left[\omega_D(q^\star)-\omega_{D^\star}(k)\right]^2
-(q^\star)^2-k^2-2q^\star k x \,,
\\
u_2(k,q^\star,x;s)
&=
\left[\omega_{D^\star}(q^\star)-\omega_D(k)\right]^2
-(q^\star)^2-k^2-2q^\star k x \,,
\end{align}
The OPE cuts satisfy the following conditions,
\begin{align}
    u_1(k,q^\star,x;s)&=m_\pi^2 \,, \\
    u_2(k,q^\star,x;s)&=m_\pi^2 \,.
    \label{eq:ope_cut_condition_u}
\end{align}
One solves for the values of $k$ that satisfy this condition, and this parameterizes the cut along the $k$ complex plane. Doing this, one finds, 
\begin{align}
    k_{1,\pm}(q^\star,x)
    &=
    \frac{-A_\pi \,q^\star x
    \pm \omega_D(q^\star)\,\sqrt{\Delta_D}}
    {2\beta_{D,x}}\,,
    \label{eq:ope_k_roots_u1}
\\
    k_{2,\pm}(q^\star,x)
    &=
    \frac{-A_\pi \,q^\star x
    \pm \omega_{D^\star}(q^\star)\,\sqrt{\Delta_{D^\star}}}
    {2\beta_{D^\star,x}}\,,
    \label{eq:ope_k_roots_u2}
\end{align}
where $A_\pi =m_D^2+m_{D^\star}^2-m_\pi^2$, and
\begin{align}
    \beta_{a,x}&=m_a^2+q^{\star 2}(1-x^2)\,,
    \\
    \Delta_D &=
    A_\pi^2
    -(2 m_Dm_{D^\star})^2
    +(2m_{D^\star}q^\star)^2\,(x^2-1)\,,
    \\
    \Delta_{D^\star}&=
    A_\pi^2
    -(2 m_Dm_{D^\star})^2
    +(2m_D q^\star)^2(x^2-1) \,.
\end{align}
The orange curves in Figs.~\ref{fig:ope_qk_reference}, \ref{fig:contour_v2}, and \ref{fig:ope_nonanalyticity_domain} correspond to $ k_{1,\pm}$.
Since $m_D$ and $m_{D^\star}$ are nearly degenerate, the two cuts lie near one another.

For us to define our contour of integration, it is necessary to understand at which energy the off-shell OPE cut first intersects the straight real-$k$ integration contour.
This can be found directly from Eq.~\eqref{eq:ope_cut_condition_u}.
Below threshold, $q^\star=i\kappa$, and for real $k$ the imaginary part of $u_{1,2}$ is proportional to $\kappa k x$.
Thus, the first overlap with the real axis occurs at $k=0$.
For the first OPE, this implies,
\begin{align}
u_1(0,q^\star,x;s)
&=
m_D^2+m_{D^\star}^2
-2m_{D^\star}\omega_D(q^\star)
=
m_\pi^2 .
\\
\Rightarrow\omega_D(q^\star_{c,1})
&=
\frac{A_\pi}{2m_{D^\star}}
\equiv \sqrt{q^{\star2}_{c,1}+m_{D}^2},
\\
\Rightarrow q^{\star2}_{c,1}
&=
\left(\frac{A_\pi}{2m_{D^\star}}\right)^2-m_D^2 \,.
\end{align}
Similarly, the second OPE contribution gives us, 
\begin{align}
u_2(0,q^\star,x;s)
&=
m_D^2+m_{D^\star}^2
-2m_D\omega_{D^\star}(q^\star)
=
m_\pi^2 
\\
\Rightarrow
\omega_{D^\star}(q^\star_{c,2})
&=
\frac{A_\pi}{2m_D}
\equiv \sqrt{q^{\star2}_{c,2}+m_{D^\star}^2},
\\
\Rightarrow
q^{\star2}_{c,2}
&=
\left(\frac{A_\pi}{2m_D}\right)^2-m_{D^\star}^2 \,.
\end{align}
Using the hadron masses from this work, namely $m_\pi = 391~{\rm MeV}$, $m_D = 1886~{\rm MeV}$ and $m_{D^\star} = 2010~{\rm MeV}$, it is easy to see that both values of $q^{\star2}_{c}$ are negative. 
Putting these two conditions together, we find the energies at which we encounter the circular cut in momentum space to be
\begin{align}
E_{c,1}
&=
\frac{A_\pi}{2m_{D^\star}}
+
\sqrt{m_{D^\star}^2+q^{\star2}_{c,1}}\,.
\\
E_{c,2}
&=
\sqrt{m_D^2+q^{\star 2}_{c,2}}
+
\frac{A_\pi}{2m_D}\,.
\end{align}
For our masses, these correspond to $a_tE_{c,1} \approx 0.676$ and $a_tE_{c,2} \approx 0.674 $. 

We find that for the majority of fits, the pole lies sufficiently far above these kinematic points, and as a result a straight contour integration along the real axis is sufficient.
As discussed below, for some of the parametrizations the pole requires a contour deformation to avoid the cut from the OPE overlapping with the contour of integration. 

The more stringent check is shown in Fig.~\ref{fig:ope_nonanalyticity_domain}, where one maps the possible OPE singularities generated by varying the intermediate momentum along the contour and the angular variable in the OPE denominator.
The blue curves delimit the corresponding non-analyticity domain.
The absence of a pinch between these curves and the chosen contour shows that the on-shell point lies in the analytic domain connected to the integration contour.

The cutoff function $H$, Eq.~\eqref{eq:Hfunc}, is analytic in the momenta entering the integral equations, but we nevertheless checked it in the same complex momentum region.  

For the $\hat\alpha=50$ parametrizations, the pole lies farther below threshold, and the straight contour can overlap the OPE non-analyticity domain.
For these cases, we use a contour inspired by the construction of Ref.~\cite{Dawid:2023jrj}, which is deformed into the complex plane as shown in Fig.~\ref{fig:contour_v2}.
The nodes are chosen so that the contour passes between the two OPE-cut branches and returns to the real axis before the cutoff.
The cuts themselves are still defined by $\zeta(k,q)=x$ with $x\in[-1,1]$, and in the diagnostic, the non-analyticity curves are generated with one momentum on the actual deformed contour.
This figure confirms that the on-shell point is indeed in the domain of analyticity.

\begin{figure}[htbp]
    \centering
    \includegraphics[width=.55\textwidth]{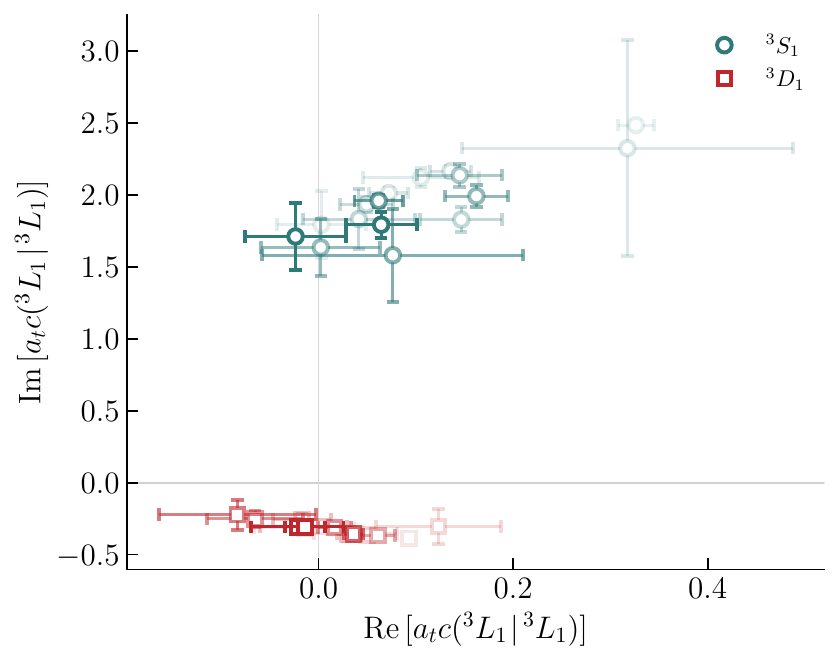}
     \caption{
     Shown are the complex-valued couplings at the $T_{cc}^+$ pole for all parametrizations considered.
     Blue circles show $c_{\SLJ{3}{S}{1}}$ and red squares show $c_{\SLJ{3}{D}{1}}$ couplings.
     The opacity follows the same relative-AIC convention used for the pole locations in Fig.~\ref{fig:pole}.
     }
    \label{fig:residues}
\end{figure}

After these contour checks, the second-sheet amplitude is obtained using Eq.~\eqref{eq:sheetII}.
Poles are found by solving
\begin{equation}
    \det\!\left[(\Mc^{\rm I})^{-1}+2i\rho\right]=0
\end{equation}
at complex values of $E^\star$, where we have used the fact that $\Mc^{II}=\left[\left(\Mc^{I}\right)^{-1}+2i\rho\right]^{-1}$.  
For each parametrization, we determine the complex pole residue using
\begin{equation}
    \Mc^{II}_{\ell'\ell} \sim -\frac{c(\SLJ{3}{\ell'}{J}) c(\SLJ{3}{\ell}{J})}{s-s_{\rm pole}}\,.
\end{equation}
The extracted couplings are shown in Fig.~\ref{fig:residues}.
The resulting pole positions and couplings are summarized in Table~\ref{tab:pole_couplings}.

\begin{table*}[htbp]
    \centering
    \setlength{\tabcolsep}{0.75em}{\renewcommand{\arraystretch}{1.75}{%
    \begin{tabular}{| l c c || c c c c | c c }
    \hline
    $\Kpar_0$ form & $\hat\alpha$ & $a_t^2 p_{\rm cut}^2$ & $a_tE_{\rm pole}$ & $a_t c(\SLJ{3}{S}{1})$ & $a_t c(\SLJ{3}{D}{1})$ & $\chi^2/N_\mathrm{dof}$  \\
    \hline\hline
    $\Kpar_0$ Poly \#3 & 170 & 0.03467 & $0.682(2)-i\,0.005(1)$ & $1.7(2)\,e^{i\,1.01(2)\frac{\pi}{2}}$ & $0.31(4)\,e^{-i\,1.05(9)\frac{\pi}{2}}$ & $\frac{27.88}{36-5}=0.90$ \\
    $\Kpar_0$ Poly \#3 & 110 & 0.04623 & $0.6804(8)-i\,0.0046(4)$ & $1.79(9)\,e^{i\,0.98(1)\frac{\pi}{2}}$ & $0.31(2)\,e^{-i\,1.03(4)\frac{\pi}{2}}$ & $\frac{27.98}{36-5}=0.90$ \\
    $\Kpar_0$ Poly \#2 & 170 & 0.03467 & $0.6798(5)-i\,0.0058(2)$ & $1.96(4)\,e^{i\,0.980(8)\frac{\pi}{2}}$ & $0.36(1)\,e^{-i\,0.94(2)\frac{\pi}{2}}$ & $\frac{30.54}{36-4}=0.95$ \\
    $\Kpar_0^{-1}$ Poly \#5 & 110 & 0.04623 & $0.682(2)-i\,0.003(2)$ & $1.6(3)\,e^{i\,0.97(7)\frac{\pi}{2}}$ & $0.27(5)\,e^{-i\,1.3(3)\frac{\pi}{2}}$ & $\frac{28.90}{36-5}=0.93$ \\
    $\Kpar_0$ Poly \#1 & 50 & 0.08090 & $0.678(1)-i\,0.0047(3)$ & $2.00(8)\,e^{i\,0.949(9)\frac{\pi}{2}}$ & $0.31(2)\,e^{-i\,0.97(3)\frac{\pi}{2}}$ & $\frac{31.19}{36-4}=0.97$ \\
    $\Kpar_0^{-1}$ Poly \#5 & 170 & 0.03467 & $0.682(1)-i\,0.004(1)$ & $1.6(2)\,e^{i\,1.00(3)\frac{\pi}{2}}$ & $0.27(4)\,e^{-i\,1.2(2)\frac{\pi}{2}}$ & $\frac{29.26}{36-5}=0.94$ \\
    $\Kpar_0$ Poly \#2 & 110 & 0.04623 & $0.678(1)-i\,0.0062(2)$ & $2.14(8)\,e^{i\,0.96(1)\frac{\pi}{2}}$ & $0.37(2)\,e^{-i\,0.90(2)\frac{\pi}{2}}$ & $\frac{31.74}{36-4}=0.99$ \\
    $\Kpar_0$ Poly \#1 & 110 & 0.04623 & $0.680(2)-i\,0.005(1)$ & $1.8(2)\,e^{i\,0.99(2)\frac{\pi}{2}}$ & $0.31(4)\,e^{-i\,1.04(8)\frac{\pi}{2}}$ & $\frac{32.34}{36-4}=1.01$ \\
    $\Kpar_0$ Poly \#3 & 50 & 0.08090 & $0.678(1)-i\,0.0043(3)$ & $1.83(9)\,e^{i\,0.95(1)\frac{\pi}{2}}$ & $0.26(3)\,e^{-i\,1.05(8)\frac{\pi}{2}}$ & $\frac{30.36}{36-5}=0.98$ \\
    $\Kpar_0^{-1}$ Poly \#6 & 170 & 0.03470 & $0.6801(6)-i\,0.0058(2)$ & $1.94(5)\,e^{i\,0.984(9)\frac{\pi}{2}}$ & $0.36(2)\,e^{-i\,0.95(2)\frac{\pi}{2}}$ & $\frac{30.54}{36-5}=0.99$ \\
    $\Kpar_0$ Poly \#4 & 110 & 0.04623 & $0.6776(6)-i\,0.0060(2)$ & $2.17(4)\,e^{i\,0.960(5)\frac{\pi}{2}}$ & $0.37(1)\,e^{-i\,0.91(1)\frac{\pi}{2}}$ & $\frac{35.02}{36-3}=1.06$ \\
    $\Kpar_0$ Poly \#4 & 170 & 0.03467 & $0.6793(5)-i\,0.0058(2)$ & $2.01(4)\,e^{i\,0.977(6)\frac{\pi}{2}}$ & $0.361(9)\,e^{-i\,0.94(1)\frac{\pi}{2}}$ & $\frac{35.30}{36-3}=1.07$ \\
    $\Kpar_0$ Poly \#2 & 50 & 0.08090 & $0.670(3)-i\,0.0065(4)$ & $2.3(8)\,e^{i\,0.94(8)\frac{\pi}{2}}$ & $0.3(1)\,e^{-i\,0.7(3)\frac{\pi}{2}}$ & $\frac{33.37}{36-4}=1.04$ \\
    $\Kpar_0^{-1}$ Poly \#6 & 110 & 0.04623 & $0.678(1)-i\,0.0058(5)$ & $2.12(7)\,e^{i\,0.97(2)\frac{\pi}{2}}$ & $0.36(4)\,e^{-i\,0.93(7)\frac{\pi}{2}}$ & $\frac{31.68}{36-5}=1.02$ \\
    $\Kpar_0$ Poly \#1 & 170 & 0.03467 & $0.681(1)-i\,0.005(1)$ & $1.8(2)\,e^{i\,1.00(2)\frac{\pi}{2}}$ & $0.32(5)\,e^{-i\,1.0(1)\frac{\pi}{2}}$ & $\frac{33.93}{36-4}=1.06$ \\
    $\Kpar_0$ Poly \#4 & 50 & 0.10979 & $0.6723(6)-i\,0.00621(5)$ & $2.51(3)\,e^{i\,0.917(4)\frac{\pi}{2}}$ & $0.398(7)\,e^{-i\,0.850(6)\frac{\pi}{2}}$ & $\frac{36.20}{36-3}=1.10$ \\
    \hline
    \end{tabular}
    }}
    \caption{
    The $T_{cc}^+$ pole positions and couplings for the parametrizations entering the systematic variations, defined by the form of $\Kpar_0$ and values of $\hat\alpha$ and $\pcut^2$.
    The results are ordered by decreasing AIC weight, and the number following ``$\#$'' corresponds to the labeling in Table~\ref{tab:parvar}.
    }
    \label{tab:pole_couplings}
\end{table*}

\end{document}